\documentclass[preprint,floats,aps,12pt]{revtex4}
\usepackage{times} 
\usepackage{amssymb,amsmath}
\usepackage{prelim2e}\usepackage[none,bottom]{draftcopy}
\draftcopyName{UT preprint / }{1.2} %ADAPT TEXT
\ifx\pdfoutput\undefined
\usepackage[usenames,dvips]{color}
\usepackage{graphicx}     % Include figure files
\else
\usepackage[usenames,dvipsnames]{color}
\usepackage{epstopdf}
\usepackage[pdftex]{graphicx}
\usepackage[pdftex]{hyperref}
\hypersetup{a4paper,
    pdftitle={Manuscript on gradient dynamics for surfactant-covered films} %ADAPT TEXT 
    pdfauthor={Uwe Thiele},  %ADAPT TEXT 
pdfproducer={lateX},
pdfview=FitV,       % FitH
pdfstartview=FitB,
linkcolor=blue,     % links to same page
citecolor=blue,     % citations
pagecolor=blue,     % links to same page
urlcolor=red,      % links to URLs
breaklinks=true,    % links may be split onto 2 lines
colorlinks=true,
citebordercolor=0 0 0,  % color for \cite
filebordercolor=0 0 0,
linkbordercolor=0 0 0,
menubordercolor=0 0 0,
pagebordercolor=0 0 0,
urlbordercolor=0 0 0,
pdfhighlight=/I,
pdfborder=0 0 0,   % no box around links
backref=false,
pagebackref=false,
bookmarks=true,
bookmarksopen=true,
bookmarksnumbered=true
}
\fi
\ifx\pdfoutput\undefined
\DeclareGraphicsExtensions{.eps}
\else
\DeclareGraphicsExtensions{.jpg, .pdf, .png, .tif}
\fi
\graphicspath{{.},{../Ps/},{./Fig/},{$HOME/Home/Obj/Sitdrop/Ps/}} %$ %ADAPT DIRECTORY
\usepackage[usenames,dvipsnames]{color}
\usepackage[normalem]{ulem}
\renewcommand{\vec}[1]{\mathbf{#1}}
\newcommand{\vecg}[1]{\boldsymbol{#1}}

\newcommand{\no}[1]{}
\newcommand{\mylab}[1]{\label{#1}}
\begin{document}
%
%----------------------------------------------------------------%
\title{Gradient dynamics models for liquid films with soluble surfactant}
\author{Uwe Thiele}
\email{u.thiele@uni-muenster.de}
\homepage{http://www.uwethiele.de}
\affiliation{Institut f\"ur Theoretische Physik, Westf\"alische
 Wilhelms-Universit\"at M\"unster, Wilhelm Klemm Str.\ 9, D-48149 M\"unster, Germany}
\affiliation{Center of Nonlinear Science (CeNoS), Westf{\"a}lische Wilhelms Universit\"at M\"unster, Corrensstr.\ 2, 48149 M\"unster, Germany}
%\affiliation{Center for Multiscale Theory and Computation (CMTC),
%  University of M\"unster, Corrensstr.\ 40, 48149 M\"unster, Germany}
\author{A. J. Archer}
\affiliation{Department of Mathematical Sciences, Loughborough University,
Loughborough, Leicestershire, LE11 3TU, UK}
\author{L. M. Pismen}
\affiliation{Department of Chemical Engineering, Technion – Israel Institute of Technology, Haifa 32000, Israel}
% \affiliation{affil}
%
\begin{abstract}
  In this paper we propose equations of motion for the dynamics of liquid
  films of surfactant suspensions that consist of a general gradient
  dynamics framework based on an underlying energy functional.
  This extends the gradient dynamics approach to dissipative
  non-equilibrium thin film systems with several variables, and casts
  their dynamic equations into a form that reproduces Onsager's
  reciprocity relations. We first discuss the general form of
  gradient dynamics models for an arbitrary number of fields and
  discuss simple well-known examples with one or two fields. Next, we
  develop the gradient dynamics (three field) model for a thin liquid film
  covered by soluble surfactant and discuss how it automatically
  results in consistent convective (driven by pressure gradients,
  Marangoni forces and Korteweg stresses), diffusive,
  adsorption/desorption, and evaporation fluxes.
  We then show that in the dilute limit, the model reduces to the
  well-known hydrodynamic form that includes Marangoni fluxes due to a
  linear equation of state. In this case the energy functional
  incorporates wetting energy, surface energy of the free interface
  (constant contribution plus an entropic term) and bulk mixing
  entropy. Subsequently, as an example, we show how various extensions of
  the energy functional result in consistent dynamical models that
  account for nonlinear equations of state, concentration-dependent
  wettability and surfactant and film bulk decomposition phase
  transitions. We conclude with a discussion of further possible
  extensions towards systems with micelles, surfactant adsorption at
  the solid substrate and bioactive behaviour.

\end{abstract}
%
%\begin{keyword} 
%Sliding drops \sep Heterogeneous substrates \sep Pinning and depinning
%\pacs{
%68.15.+e, % Thin films: Liquid thin films
%47.20.Ky  % Fluid dynamics: Nonlinearity (including bifurcation theory)
%47.55.Dz  % Drops and bubbles 
%68.08.-p  % Liquid-solid interfaces
%}
%\end{keyword} 
%
\maketitle
%
%\received{6.5.2002}
%
%----------------------------------------------------------------%
%
%%%%%%%%%%%%%%%%%%%%%%%%%%%%%%%%%%%%%%%%%%%%%%%%%%%%%%%%%%%%%%%%%%%%%%%%%%%%%%%
\section{Introduction} \mylab{sec:intro}
%%%%%%%%%%%%%%%%%%%%%%%%%%%%%%%%%%%%%%%%%%%%%%%%%%%%%%%%%%%%%%%%%%%%%%%%%%%%%%%

Onsager's evolution equations \cite{Onsa1931pr,Onsa1931prb}, based on
the principle of detailed balance embedded in Onsager's reciprocity
relations, became a key tool for understanding the relaxational
approach to equilibrium in a variety of physical processes. More
recently, Doi \cite{Doi2011jpcm} extended the range of this approach
to processes in macroscopic soft matter systems, such as the swelling
of gels and the dynamics of liquid crystals. It is less obvious that a
similar approach can also be applied to processes out of equilibrium in
spatially extended open systems.  A well known example is the dynamics
of single layer thin films in the long-wave (or lubrication) approximation
\cite{OrDB1997rmp,CrMa2009rmp} where a single variable -- the layer
thickness -- is sufficient for a description of the system.  In this
case, it is not a-priori obvious that an energy functional of thermodynamic origin
exists for the system. Nevertheless, as noticed by Mitlin \cite{Mitl1993jcis} for
dewetting films and by Rosenau and Oron for thin films heated from
below \cite{OrRo1992jpif}, the dynamic equation for the layer thickness
$h$ can be cast into a gradient dynamics form
\begin{equation}
\partial_t h \,=\,
\nabla\cdot\left[Q^{\mathrm{c}}\nabla\frac{\delta
    \mathcal{F}}{\delta h}\right]
- Q^{\mathrm{nc}}\left(\frac{\delta \mathcal{F}}{\delta h} -p_\mathrm{vap}\right),
\mylab{eq:onefield:gov}
\end{equation}
showing that the evolution can be derived from a certain ``energy''
functional $\mathcal{F}[h]$. $p_\mathrm{vap}$ is the pressure of the
vapour phase that may instead be incorporated into
$\mathcal{F}$. Here and in the following $\partial_t$ denotes partial
time derivatives and $\nabla$ is the two-dimensional (2D) spatial gradient operator.
Eq.~\eqref{eq:onefield:gov} is the general form in which the dynamics has both a conserved and
a non-conserved contributions with mobilities $Q^{\mathrm{c}}(h)\ge0$
and $Q^{\mathrm{nc}}(h)\ge0$, respectively \cite{Thie2010jpcm}.

The usual procedure of irreversible thermodynamics is thereby
reversed: first comes a dynamic equation obtained through a series of
simplifications, and then a suitable functional is assigned, ensuring
a dissipative evolution toward a minimum of this energy. However, in the
case of dewetting the energy functional is the ``interface
Hamiltonian'' that is obtained via a systematic coarse-graining
procedure from the microscale interaction energies
\cite{BEIM2009rmp}. Sometimes, even systems that are permanently out
of equilibrium can be accommodated, as in the case of sliding droplets
on an infinitely extended incline, where the correct thin film model
can be brought into the form of a gradient dynamics with an underlying
energy functional that includes potential energy \cite{EWGT2016arxiv}.

Besides long-wave thin film equations, other examples of one-field
gradient dynamics are the Cahn-Hilliard equation describing the
demixing of a binary mixture, i.e., a purely conserved dynamics
($Q^{\mathrm{nc}}=0$) \cite{CaHi1958jcp,Cahn1965jcp,Lang1992} and the
Allen-Cahn equation that models, for instance, the purely
non-conserved dynamics ($Q^{\mathrm{c}}=0$) of the Ising model in the
mean field continuum limit \cite{Lang1992}. In general, equations of
the form (\ref{eq:onefield:gov}) are ubiquitous. They appear with
various choices of $\mathcal{F}$, not only in the context of the dynamics of
films of non-volatile and volatile liquids on solid substrates
\cite{OrDB1997rmp,Mitl1993jcis,PiPo2000pre,Thie2014acis}, but also as
evolution equations for surface profiles in epitaxial growth
\cite{SpVD1991prl,GoDN1999pre,GLSD2004prb,Vved2004jpcm,Thie2010jpcm},
and, indeed, as models of one-component lipid bilayer adhesion
dynamics \cite{GaCB1993jcis}. Another field of application is in
dynamical density functional theory (DDFT), describing the dynamics
of the density distribution of colloidal particles
\cite{MaTa1999jcp,MaTa2000jpm,ArEv2004jcp,ArRa2004jpag}.

Furthermore, many hydrodynamic
two- and more-field long-wave models were developed that describe,
e.g., the evolution of multilayer films, films of mixtures or
surfactant-covered films \cite{CrMa2009rmp}. Normally, they are not
written in the gradient dynamics form. However, recently, the gradient
dynamics approach was extended to several two-field models, namely,
for the dewetting of two-layer films \cite{PBMT2004pre,JHKP2013sjam},
for the coupled decomposition and dewetting of a film of a binary
mixture \cite{Thie2011epjst,ThTL2013prl} and for the evolution of a
layer of insoluble surfactant on a thin liquid film
\cite{ThAP2012pf}. In all these cases, energies with a clear physical
meaning can be given that may also be obtained via the coarse-graining
procedures of statistical physics. Note though that
the description of a thin two layer-film heated from below cannot be
brought into the Onsager form \cite{PBMT2005jcp}, marking the single
layer case as a fortuitous `accident'. Nonetheless, certain out of equilibrium phenomena
can be described via the addition of appropriate potential energies
to the energy functional or, as in the case of dip coating and
Langmuir--Blodgett transfer through ``comoving frame terms'' that
account for a moving substrate that is withdrawn from a bath
\cite{WTGK2015mmnp}.  Similar two-field gradient dynamics models exist
for the dynamics of membranes \cite{SaGo2007pre,HiKA2012pre} or as
DDFTs for mixtures \cite{ArRT2010pre,archer2005jpcm}.

The aim of this paper is to extend the gradient dynamics
approach to describe the non-equilibrium dissipative dynamics of
thin film systems with several
variables, and to cast the dynamic equations into a form that
reproduces Onsager's reciprocity relations. A further aim is to
incorporate interphase exchange processes, such as evaporation and
surfactant dissolution to derive equations combining conserved
(Cahn--Hilliard-type) and non-conserved (reaction-diffusion, or
Allen--Cahn-type) terms. In doing so, several limitations of the known
two-field models are alleviated. The particular example treated in
detail is a thin liquid film that is covered by a soluble surfactant
and rests on a solid substrate. The gradient dynamics model then
describes the coupled evolution of the film height profile, the amount
of surfactant within the film and the surface concentration dynamics
(three field) model for the case of a thin liquid film covered by a
soluble surfactant as sketched in Fig.~\ref{fig:sketch}.

\begin{figure}[t]
\includegraphics[width=0.49\hsize]{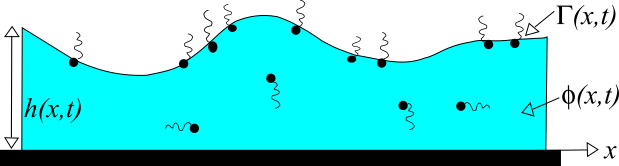}
\caption{Sketch of the system under consideration. It consists of a
film of liquid on a surface of thickness $h(x,t)$, that varies with
location on the surface
$x$ and with time $t$. On the liquid film free surface are surfactant
molecules, with local density $\Gamma(x,t)$. The surfactant
molecules have some solubility in the liquid and the local concentration
within the body of the liquid is $\phi(x,t)$. We assume that $\phi$
does not vary
vertically and only varies horizontally and with $t$. This is
equivalent to treating $\phi$ as a height-averaged concentration.
Over time there is exchange of surfactant molecules between the
surface of the liquid and the bulk. There can also be condensation
or evaporation of the liquid to vapour in the air above.}
\mylab{fig:sketch}
\end{figure}

This paper is structured as follows: In the following
section~\ref{sec:multifield} we discuss the general form of gradient
dynamics models, first, for an arbitrary number of fields in
section~\ref{sec:onefield:general} and then in
section~\ref{sec:examples} we write the diffusion equation and the thin
film equation as gradient dynamics and discuss known two-field models.
Next, in section~\ref{sec:surfsol:gdm} we develop the gradient
dynamics (three field) model for the case of a thin liquid film
covered by a soluble surfactant and discuss in
section~\ref{sec:surfsol-special} special cases and extensions. We
draw our conclusions in
section~\ref{sec:conc}. Appendices~\ref{sec:lwvar} and \ref{app:variations} 
clarify an issue in the comparison
of hydrodynamic long-wave approach and the present variational
approach and give the variations of the energy functional in the most
general case covered by the present work, respectively.

%%%%%%%%%%%%%%%%%%%%%%%%%%%%%%%%%%%%%%%%%%%%%%%%%%%%%%%%%%%%%%%%%%%%%%%%%%%%%%%
\section{General $N$-field model and known applications}
\mylab{sec:multifield}
%%%%%%%%%%%%%%%%%%%%%%%%%%%%%%%%%%%%%%%%%%%%%%%%%%%%%%%%%%%%%%%%%%%%%%%%%%%%%%%
%%%%%%%%%%%%%%%%%%%%%%%%%%%%%%%%%%%%%%%%%%%%%%%%%%%%%%%%%%%%%%%%%%%%%%%%%%%%%%%
\subsection{General model}
\mylab{sec:onefield:general}
%%%%%%%%%%%%%%%%%%%%%%%%%%%%%%%%%%%%%%%%%%%%%%%%%%%%%%%%%%%%%%%%%%%%%%%%%%%%%%%
%
The dynamics of a spatially extended system may be characterised by the
coupled evolution of $N$ scalar state variable fields (order
parameter fields) $\mathbf{u} = (u_1,u_2,..., u_n)^\mathrm{T}$. Not
too far from equilibrium, the dynamics is governed by a single
equilibrium free energy functional $\mathcal{F}[\mathbf{u}]$,
i.e., it is a gradient dynamics.
Using Einstein's index notation that presumes summation over repeated indices, the coupled evolution equations read 
\begin{equation}
 \partial_t u_a=\nabla_\alpha \left[Q^\mathrm{c}_{ab}
   \nabla_\alpha\frac{\delta \mathcal{F}}{\delta u_b}\right]
- Q^\mathrm{nc}_{ab} \frac{\delta \mathcal{F}}{\delta u_b}
\mylab{eq:nn3}
\end{equation}
where $\alpha=1,2,\dots,d$ refers to spatial coordinates and $a, b =
1,\dots,n$ refer to the different order parameter fields that might
have a conserved, or non-conserved, or mixed dynamics. Here, $Q^\mathrm{c}_{ab}(\mathbf{u})$ and $Q^\mathrm{nc}_{ab}(\mathbf{u})$ represent $n \times n$
dimensional \textit{positive definite} and \textit{symmetric} mobility
matrices for the conserved and non-conserved parts of the dynamics,
respectively. The mobilities $Q^\mathrm{c}_{ab}$ govern the fluxes
$j_a=-Q^\mathrm{c}_{ab}\nabla({\delta\mathcal{F}}/{\delta u_b})$ of
the conserved part of the dynamics for all order parameters $u_a$.
These are given as linear combinations of the influences of all
thermodynamic forces $-\nabla({\delta \mathcal{F}}/{\delta 
u_b})$, i.e.\ are linear in the thermodynamic forces. In contrast, the
coefficients $Q^\mathrm{nc}_{ab}$ give the transition rates between
fields and are also linear combinations of the thermodynamic potentials
${\delta \mathcal{F}}/{\delta u_a}$.

It is straightforward to show that the free energy $\mathcal{F}[u_1,\dots,u_n] $
is a Lyapunov functional, i.e., it monotonically decreases in time:
\begin{align}
 \frac{\mathrm{d}}{\mathrm{d}t}\mathcal{F}[u_1,\dots,u_n]&=
\int\limits_{\Omega} \frac{\delta \mathcal{F}}{\delta u_a}
\frac{\partial u_a}{\partial t}~\mathrm{d}^dx\nonumber \\
&=\int\limits_{\Omega}\frac{\delta \mathcal{F}}{\delta
  u_a}\nabla_\alpha \left[Q^\mathrm{c}_{ab}
  \nabla_\alpha\frac{\delta \mathcal{F}}{\delta u_b}\right] ~\mathrm{d}^dx
- \int\limits_{\Omega} \frac{\delta \mathcal{F}}{\delta u_a}
Q^\mathrm{nc}_{ab}\frac{\delta \mathcal{F}}{\delta u_b}~\mathrm{d}^dx
\nonumber \\
&=-\int\limits_{\Omega} \left(\nabla_\alpha\frac{\delta \mathcal{F}}{\delta
    u_a}\right)  Q^\mathrm{c}_{ab}  \left(\nabla_\alpha\frac{\delta
    \mathcal{F}}{\delta u_b}\right)~\mathrm{d}^dx
- \int\limits_{\Omega} \frac{\delta \mathcal{F}}{\delta u_a}
Q^\mathrm{nc}_{ab}\frac{\delta \mathcal{F}}{\delta u_b}~\mathrm{d}^dx
\leq 0.
\end{align}
where $\Omega$ is the domain in which the system is defined.
 Above we used Eq.~(\ref{eq:nn3}) and partial integration, assuming
 periodic or no-flux boundary conditions.

A further advantage of the general formulation is the ease with which
one may change the choice of variables $u_a$. If new order
parameter fields $\tilde u_a $ are introduced via a linear
transformation $\tilde u_a = R_{ab}u_b$ the
kinetic equations for the new fields are
\begin{equation}
 \partial_t \tilde u_a=\nabla_\alpha \left[\widetilde Q^\mathrm{c}_{a b}(\tilde u_1,\dots,\tilde u_n) \nabla_\alpha\frac{\delta
     \mathcal{F}}{\delta \tilde u_b}\right]
\,-\, \widetilde Q^\mathrm{nc}_{a b}(\tilde u_1,\dots,\tilde u_n)
\frac{\delta \mathcal{F}}{\delta\tilde  u_b}
\mylab{eq:trans-gen}
\end{equation}
with $\widetilde Q^\mathrm{i}_{a b}=R_{a
  d}Q^\mathrm{i}_{de}R_{be}$
($\mathrm{i}=\mathrm{c}, \mathrm{nc}$) where we take into account that 
$\delta \mathcal{F}/\delta u_a=R_{b a}\delta \mathcal{F}/\delta
\tilde u_b$. For two conserved fields, similar relations were already given in
Refs.~\cite{XuTQ2015jpcm,WTGK2015mmnp}.

Up to here, we have not specified the free energy
$\mathcal{F}[u_1,\dots,u_n]$ that can, in principle, be an arbitrary
functional of the order parameter fields. If $\mathcal{F}$ is a multiple
integral, Eq.~(\ref{eq:nn3}) becomes a system of
integro-differential equations,
as is often the case in DDFT for
colloids \cite{ArEv2004jcp}. However, often the kernel is expanded in
derivatives of the order parameter fields and Eq.~(\ref{eq:nn3})
corresponds to a system of partial differential equations. Examples
are Phase Field Crystal (PFC) models \cite{ELWG2012ap} and membrane
models \cite{HiKA2009c,HiKA2012pre}
where the highest order terms in the energy are $\sim(\Delta u)^2$. Here we restrict our 
attention to a lower order and only consider models where the highest 
order terms are $\sim(\nabla u)^2$. Then the general form is
\begin{equation}
 \mathcal{F}[u_1,\dots,u_n]=\int\limits_{\Omega} \left[ \frac{1}{2}
   (\nabla_\alpha u_a)~ \Sigma_{a b}  ~(\nabla_\alpha u_b) 
+ f(u_1,\dots,u_n)\right] ~\mathrm{d}^dx, 
\mylab{eq:nfield}
\end{equation}
where we have introduced in the free energy a symmetric $n \times n$
dimensional gradient interaction matrix $\Sigma_{a b}$ that,
in principle, may itself also depend on $\mathbf{u}$.
The integrand may also contain metric factors (see below).

Before we come in Section~\ref{sec:surfsol:gdm} to the case of liquid
films that are covered with a soluble surfactant, we briefly review in
Section~\ref{sec:examples} some basic examples where only one or two
order parameter fields are involved.
%
%%%%%%%%%%%%%%%%%%%%%%%%%%%%%%%%%%%%%%%%%%%%%%%%%%%%%%%%%%%%%%%%%%%%%%%%%%%%%%%
\subsection{Specific known examples of gradient dynamics}
\mylab{sec:examples}
%%%%%%%%%%%%%%%%%%%%%%%%%%%%%%%%%%%%%%%%%%%%%%%%%%%%%%%%%%%%%%%%%%%%%%%%%%%%%%%
%%%%%%%%%%%%%%%%%%%%%%%%%%%%%%%%%%%%%%%%%%%%%%%%%%%%%%%%%%%%%%%%%%%%%%%%%%%%%%%
\subsubsection{Diffusion equation}
\mylab{sec:onefield:diff}
%%%%%%%%%%%%%%%%%%%%%%%%%%%%%%%%%%%%%%%%%%%%%%%%%%%%%%%%%%%%%%%%%%%%%%%%%%%%%%%
%
In the dilute limit, the diffusion of a species with part-per-volume concentration $c$ in a quiescent carrier medium can
be represented as the conserved gradient dynamics
\begin{equation}
\partial_t  c\,=\,
\nabla\cdot\left[Q^\mathrm{c}_{cc}\nabla\frac{\delta \mathcal{F}}{\delta c}\right],
\mylab{eq:trans-diff}
\end{equation}
with the purely entropic Helmholtz free energy functional 
\begin{equation}
\mathcal{F}[c]\,=\,\frac{kT}{l^3}\int c [\ln c - 1] \,dV,
\mylab{eq:en-diff}
\end{equation}
where $k$ is Boltzmann's constant, $T$ is the temperature and $l$ is a
molecular length scale. The
mobility function in Eq.~(\ref{eq:trans-diff}) is $Q^\mathrm{c}_{cc}=\widetilde{D} c$ and can be obtained via
Onsager's variational principle \cite{Doi2011jpcm,Doi2013,XuTQ2015jpcm}. Here, $\widetilde{D}$ 
is the molecular mobility. This corresponds to $\partial_t c \,=\, - \nabla\cdot j_\mathrm{diff}$
where Fick's law takes the form $j_\mathrm{diff}=-\widetilde{D}
c\nabla \mu =-D\nabla c$ with the chemical potential $\mu=\delta
\mathcal{F}/\delta c = (kT/l^3)\ln c$, i.e. $D=\widetilde{D}c\,d\mu/dc=\widetilde{D}kT/l^3$.

The equivalence of Eq.~(\ref{eq:trans-diff}) and the standard
diffusion equation has been easily shown, and now allows one to use the
advantages of the gradient dynamics form, namely, the straightforward
way to account for free energies that are not purely entropic. If, for
instance, one replaces the integrand in $\mathcal{F}[c]$ of
Eq.~(\ref{eq:en-diff}) by the sum of a double-well potential and a
squared gradient term, one obtains the Cahn-Hilliard equation
(then using a constant $Q^\mathrm{c}_{cc}$) \cite{Cahn1965jcp}.

%%%%%%%%%%%%%%%%%%%%%%%%%%%%%%%%%%%%%%%%%%%%%%%%%%%%%%%%%%%%%%%%%%%%%%%%%%%%%%%
\subsubsection{Thin films of simple liquids}
\mylab{sec:onefield:onelay}
%%%%%%%%%%%%%%%%%%%%%%%%%%%%%%%%%%%%%%%%%%%%%%%%%%%%%%%%%%%%%%%%%%%%%%%%%%%%%%%
%
As discussed above, Eq.~(\ref{eq:onefield:gov}) describes the
evolution of the height profile of a thin liquid film on a solid
substrate for non-volatile ($Q^{\mathrm{nc}}(h)=0$) or volatile
($Q^{\mathrm{nc}}(h)\ge0$) liquids. Detailed discussions of the
various physical situations treated can be found in 
\cite{OrDB1997rmp,Thie2010jpcm,Thie2014acis}. In the most basic case of
mesoscopic hydrodynamics, only the influence of capillarity and
wettability is considered. The corresponding free energy
$\mathcal{F}[h]$ is then
\begin{equation}
\mathcal{F}[h]\,=\,\int\left[\frac{\gamma}{2}(\nabla h)^2 + g(h)\right]d^2x,
\mylab{eq:en1}
\end{equation}
where $\gamma$ is the surface tension of the liquid and $g(h)$ is a
local free energy (wetting or adhesion energy, or binding potential),
related to the Derjaguin (or disjoining) pressure $\Pi(h)$ by
$\Pi=-dg(h)/dh$ \cite{Genn1985rmp}. Note, that varying sign conventions are used
  throughout the literature. For particular forms of $\Pi$, see,
e.g., Refs.~\cite{Genn1985rmp,Israelachvili2011,Mitl1993jcis,OrDB1997rmp,PiPo2000pre,KalliadasisThiele2007}.
Similar expressions are obtained as ``interface Hamiltonians'' in the
context of wetting transitions \cite{BEIM2009rmp}. Therefore
mesoscopic thin film (or two-dimensional) hydrodynamics might be considered as a gradient
dynamics on the underlying interface Hamiltonian. Note that recently
such mesoscopic wetting energies have been extracted via parameter
passing methods from different microscopic models (molecular dynamics and
density functional theory) \cite{TMTT2013jcp,HuTA2015jcp}.
Without slip at the substrate, $Q^{\mathrm{c}} \equiv Q^{\mathrm{c}}_{hh}=h^3/3\eta$, where 
$\eta$ is the dynamic viscosity. Different slip models can be
accommodated by alternative choices of $Q^{\mathrm{c}}_{hh}$ \cite{MWW2005jem}.
Although several functions $Q^{\mathrm{nc}}(h)$ are discussed in the literature for the 
case of volatile liquids (see, e.g., \cite{Thie2014acis}), often a
constant is used \cite{LyGP2002pre}).
%
%%%%%%%%%%%%%%%%%%%%%%%%%%%%%%%%%%%%%%%%%%%%%%%%%%%%%%%%%%%%%%%%%%%%%%%%%%%%%%%
\subsubsection{Two-field models}
\mylab{sec:twofield}
%%%%%%%%%%%%%%%%%%%%%%%%%%%%%%%%%%%%%%%%%%%%%%%%%%%%%%%%%%%%%%%%%%%%%%%%%%%%%%%

In the context of thin film hydrodynamics, two-field gradient dynamics
models were presented and analysed (i) for dewetting two-layer films
on solid substrates, i.e., staggered layers of two immiscible fluids
\cite{PBMT2004pre,PBMT2005jcp,JHKP2013sjam}, (ii) for decomposing and
dewetting films of a binary liquid mixture (with non-surface active
components) \cite{Thie2011epjst,ThTL2013prl}, and (iii) for the
dynamics of a liquid film that is covered by an insoluble surfactant
\cite{ThAP2012pf}. In all three cases, the model has the form
(\ref{eq:nn3}) with $a, b = 1,2$ and all
$Q^\mathrm{nc}_{a b}=0$ (purely conserved dynamics). The
conserved fields $u_1$ and $u_2$ represent in case (i) the lower layer
thickness $h_1$ and overall thickness $h_2$, respectively
\cite{PBMT2004pre,PBMT2005jcp,JHKP2013sjam} or the
lower and upper layer thickness
\cite{BCJP2013epje} (the transformation between the two formulations follows from
the discussion around Eq.~(\ref{eq:trans-gen})). In case (ii), $u_1$ and
$u_2$ represent the film height $h$ and the effective solute height
$\psi=ch$, respectively, where $c$ is the height averaged
concentration.  Finally, in case (iii), $u_1$ and $u_2$ represent the
film height $h$ and the surfactant coverage $\widetilde\Gamma$ (that
is projected on the cartesian substrate plane), respectively
\cite{ThAP2012pf}.

As already emphasised, a crucial point in cases (ii) and (iii) is the
choice of the two fields that can be varied independently of each
other. This is not the case if, e.g., film height $h$ and height
averaged concentration $c$ are used in case (ii), since then a variation
in the height for fixed particle number per substrate area implies that $c$
varies \cite{Thie2011epjst}. In case (iii), the projected coverage $\widetilde\Gamma$ has
to be used since the surfactant coverage $\Gamma $ on the free surface
and the height profile $h$ are not independent \cite{ThAP2012pf}: If
the slope of $h$ changes locally, the surface area changes and so also $\Gamma$.
Therefore, for a fixed local number of surfactant molecules, the local concentration
changes without any surfactant transport. If one uses dependent
fields, one is not able to employ the general form
(\ref{eq:nn3}). Note that in Refs.~\cite{Clar2005m,TCPM2010sm} case
(ii) has been treated employing a gradient dynamics for $h$ and
$c$. For a further comparison with the approach employed in
\cite{Thie2011epjst,ThTL2013prl}, see Ref.~\cite{Thie2014acis}. In all
three cases (i) to (iii) the underlying free energy functionals have a
clear thermodynamic significance. They may be seen as extensions of
the interface Hamiltonian for a single adsorbed layer, and the
individual terms may be obtained from equilibrium statistical
physics. As expected, the mobility matrices $\mathbf{Q}^\mathrm{c}$
are positive definite and symmetric
\cite{PBMT2004pre,BCJP2013epje,ThTL2013prl,ThAP2012pf}. All their
entries are low order polynomials in the respective fields $u_1$ and
$u_2$. In particular, in cases (ii) and (iii), one has
\begin{equation}
\mathbf{Q}^\mathrm{c}\,=\,
\,=\,\frac{1}{3\eta}\left( 
\begin{array}{cc}  
u_1^3 & qu_1^2u_2 \\[.3ex]
qu_1^2u_2 & \quad ru_1u_2^2+ 3\widetilde{D}\eta u_2
\end{array}
\right).
\mylab{eq:mob}
\end{equation}
where $\widetilde{D}$ is a respective molecular mobility related to
diffusion, and $q=r=1$ in case (ii) and $q=3/2$, $r=3$ in case (iii).
Actually, in the parametrisation of Ref.~\cite{JHKP2013sjam}, the mobility matrix
$\mathbf{Q}^\mathrm{c}$ of case (i) also agrees with case (iii) if the
diffusion term $3\widetilde{D}\eta u_2$ is replaced by
$\eta_\mathrm{r}u_2^3$ where $\eta_\mathrm{r}$ is the viscosity ratio
of the two layers.

Note in particular that cases (ii) and (iii) in the respective
low concentration limit give the known hydrodynamic thin film equations
coupled to the equation for the solute / surfactant as discussed in
detail in Refs.~\cite{Thie2011epjst} and \cite{ThAP2012pf},
respectively. It recovers also a number of other special cases and can
be employed to devise models that incorporate various energetic
cross-couplings in a thermodynamically consistent manner. Examples
include wetting energies that depend on solute or surfactant
concentration, effects of surface rigidity for surfactant covered
films, free energies of mixing/decomposition including gradient
contributions, etc. It also allows one to discuss the influence of
solutes / surfactants on evaporation.

Note that the discussion above mixes the possible extensions in cases (ii) and
(iii) that are separately discussed in \cite{Thie2011epjst} and
\cite{ThAP2012pf}, respectively. It was noted in \cite{ThTL2013prl}
that the two-field model for a film of a mixture cannot accommodate a
solutal Marangoni effect by simply incorporating a
concentration-dependent surface tension since this breaks the gradient
dynamics structure. Another disadvantage of the two-field model is that
most surfactants are soluble, a situation that cannot be treated via case
(iii). In the following, we develop a three-field
model that alleviates all the mentioned problems. 
%
%%%%%%%%%%%%%%%%%%%%%%%%%%%%%%%%%%%%%%%%%%%%%%%%%%%%%%%%%%%%%%%%%%%%%%%%%%%%%%%
\section{Soluble surfactant - gradient dynamics model} \mylab{sec:surfsol:gdm}
%%%%%%%%%%%%%%%%%%%%%%%%%%%%%%%%%%%%%%%%%%%%%%%%%%%%%%%%%%%%%%%%%%%%%%%%%%%%%%%
\subsection{Energy functional}
\mylab{sec:surfsol:energy}
We consider a thin film of liquid of thickness $h$ on a
solid substrate with a free surface that is
covered by a soluble surfactant, i.e., part of the
surfactant is dissolved in the bulk of the film and part is adsorbed at the
free surface -- see Fig.~\ref{fig:sketch}.
We neglect adsorption at the solid-liquid interface and micelle
formation but discuss in teh conclusion how they can be incorporated. The
surfactant concentration $\phi$ within the film represents a
height-averaged concentration, i.e., it is assumed that the
concentration is nearly uniform over the film layer thickness. The system
is considered in relaxational situations, i.e., the boundary
conditions do not sustain energy or mass fluxes. Therefore, we
expect the system dynamics to follow a pathway that approaches a
static equilibrium.  In the absence of evaporation and surfactant
exchange between the interface and the bulk solution, the approach to
equilibrium can be described by gradient dynamics for three
\textit{independent fields}: the film thickness $h(\mathbf{r},t)$, the
local amount of dissolved surfactant
$\psi(\mathbf{r},t)=h(\mathbf{r},t)\,\phi(\mathbf{r},t)$, and the
surfactant concentration at the interface projected onto a cartesian
reference plane $\widetilde \Gamma(\mathbf{r},t)$.  The surfactant
concentration on the interface is given by
$\Gamma= \widetilde \Gamma/\sqrt{a}$ where $a$ is the determinant of
the surface metric tensor (see below). Here $\mathbf{r}=(x,y)$
are ``horizontal'' coordinates in the substrate plane. The fields
$\phi$ and $\Gamma$ are expressed as volume fraction and area fraction
concentrations, respectively, i.e., they are both dimensionless.  As
emphasised in section~\ref{sec:twofield} for the two-field cases,
variations in $h$, $\phi$ and $\Gamma$ are not independent, whilst variations
with respect to $h$, $\psi$ and $\widetilde\Gamma$ are independent.

The general expression for the energy includes surface and bulk contributions: 
\begin{eqnarray}
\mathcal{F} &=& \mathcal{F}_\mathrm{s}+\mathcal{F}_\mathrm{b} =
 \int(\mathcal{L}_\mathrm{s} +\mathcal{L}_\mathrm{b}) dx dy, 
 \mylab{eq:ff} \\
\mathcal{L}_\mathrm{s} &=& \left[ \frac{\kappa_\mathrm{s}}{2}
  a^{\alpha\beta}(\partial_\alpha\Gamma) (\partial_\beta\Gamma) 
+ f_\mathrm{s} (\Gamma)  \right] \sqrt{a}+ g(h),
\mylab{eq:insolsurf} \\
\mathcal{L}_\mathrm{b} &=&  h \left[  \frac{\kappa}{2}|\nabla\phi|^2 +f(\phi) \right]. 
\mylab{eq:solsurf}
\end{eqnarray}
The interfacial terms in Eq.~(\ref{eq:insolsurf}) depend on the
surface metric tensor
$a_{\alpha\beta}= \delta_{\alpha\beta} + \partial_\alpha
h \partial_\beta h$
[where we exclude overhangs in order to use a Monge representation
$h(x,y)$] and its inverse $a^{\alpha\beta}$; $a$ is the determinant of
$a_{\alpha\beta}$ and determines the extension of the interface, and
$\delta_{\alpha\beta}$ is the Cartesian metric of the planar substrate
or a planar surface $h=$ const. Distinction between lower (covariant)
and upper (contravariant) indices is essential for a non-Euclidean
surface metric. The wetting potential $g(h)$ in
Eq.~(\ref{eq:insolsurf}) describes the interactions with the substrate
that determine the Derjaguin (or disjoining) pressure
$\Pi(h)=-dg(h)/dh$ (cf.~Section~\ref{sec:onefield:onelay}).  The first
terms in parentheses in Eqs.~(\ref{eq:insolsurf}) and
(\ref{eq:solsurf}) contain the interfacial and bulk rigidity
coefficients $\kappa_\mathrm{s}$ and $\kappa$, respectively, and
penalize surfactant concentration gradients.  The second terms in the
parentheses in each case takes account of molecular
interactions. $f_\mathrm{s}(\Gamma)$ contains the free energy
contribution due to the presence of surfactant molecules at the
interface. In the limit $\Gamma\to0$, then this is just the pure
liquid-vapour surface tension, i.e.\
$f_\mathrm{s}(\Gamma \to 0)=\gamma_0$, but more generally
\begin{equation}
f_\mathrm{s}(\Gamma)=\gamma_0+\frac{kT}{l_s^2}\Gamma[\ln\Gamma-1]+f_s^{ex}(\Gamma).
\mylab{eq:enfsurf0}
\end{equation}
The second term is the contribution to the free energy when the amount of surfactant on the surface is low enough that interactions between molecules are negligible and can be treated as a 2D ideal-gas. $l_s$ is a molecular length scale related to the size of the adsorbed surfactant molecules ($l_s^2$ is the area on the surface occupied by a surfactant molecule). As the surface coverage $\Gamma$ increases, then the excess free energy $f_\mathrm{s}^{ex}(\Gamma)$ gives an increasing contribution. For example, treating the surfactant on the surface via a lattice-gas approximation, one would write
\begin{equation}
f_\mathrm{s}^{ex}(\Gamma)=\frac{kT}{l_s^2}[\Gamma+(1-\Gamma)\ln(1-\Gamma)] - \frac{b}{2}\Gamma^2
\mylab{eq:enfsurf}
\end{equation}
where the first (entropic) excluded volume term comes from assuming
only one surfactant molecule can occupy a site of area $l_\mathrm{s}^2$ on the
surface and the final term is a simple mean-field term coming from the
attraction between pairs of neighbouring surfactant molecules. If the
attraction strength parameter $b>0$ is sufficiently large, then
surface phase transitions may occur. An alternative approximation
might be $f_\mathrm{s}^{ex}(\Gamma)=f_{hd}(\Gamma)-b\Gamma^2/2$, where $f_{hd}$
is the hard-disk excess free energy -- see for example the
approximations in Refs.~\cite{APER2007jcp,AIPR2008jpm}.

Similarly, the bulk free energy in Eq.~(\ref{eq:solsurf}) can be written as:
\begin{equation}
f(\phi)=\frac{kT}{l^2}\phi[\ln\phi-1]+f_b^{ex}(\phi),
\mylab{eq:enfbulk0}
\end{equation}
where $l$ is a molecular length scale related to the surfactant
molecules in solution ($l^3$ is the volume occupied by a surfactant
molecule). The simplest approximation is to assume $l_s =l$.
$f_b^{ex}(\phi)$ is the bulk excess contribution which in general may
be written as a virial expansion
$f_b^{ex}(\phi)=\sum_{i=2}^\infty c_i\phi^i$, with coefficients $c_i$
that depend on the temperature. Alternatively one may approximate,
e.g.\ by assuming a lattice-gas free energy
\begin{equation}
f_b^{ex}=\frac{kT}{l^2}[\phi+(1-\phi)\ln(1-\phi)] - \frac{b_b}{2}\phi^2,
\mylab{eq:enfbulk}
\end{equation}
where $b_b>0$ is an inter-surfactant molecule attraction strength
parameter. Or, instead one could assume
$f_b^{ex}(\phi)=f_{cs}(\phi)-b_b\phi^2/2$, where $f_{cs}(\phi)$ is the
Carnahan-Starling approximation for the hard-sphere excess free energy
\cite{HansenMcDonald2006}.  Specific cases for the excess
contributions $f_\mathrm{s}^{ex} (\Gamma) $ and $f_\mathrm{ex} (\phi)$
will be discussed below in
Section~\ref{sec:surfsol-nonconsfluxadsorp}.

\subsection{Pressures, chemical potentials and surface stress}
\mylab{sec:surfsol:press}
 The expression for pressure $p=\delta \mathcal{F}/\delta h$ is
obtained by calculating the variation of Eq.~(\ref{eq:ff}) with respect to $h$ for fixed
$\widetilde{\Gamma}, \psi$.  The variation of $\mathcal{F}_\mathrm{s}$ depends
on the surface metric and uses the relations
\begin{equation}
\delta a = a\,a^{\alpha\beta} \delta a_{\alpha\beta}
= -a\,a_{\alpha\beta} \delta a^{\alpha\beta}, \quad
g\delta  a_{\alpha\beta}=g \delta (\partial_\alpha h \partial_\beta h)
= -[ \partial_\alpha(g \partial_\beta h) + \partial_\beta(g \partial_\alpha h)]\delta h,
\mylab{eq:varh}
\end{equation}
where $g$ is an arbitrary function of the surface coordinates. 
Also note that $a^{\alpha\beta} a_{\alpha\beta}=\delta_\alpha^\alpha=2$.
As
mentioned above, $\Gamma$ changes with surface extension or
contraction, so that before the variation of $\mathcal{F}_\mathrm{s}$ is
computed one needs to replace $\Gamma=\widetilde{\Gamma}/\sqrt{a}$,
where $\widetilde{\Gamma}$ is a reference surfactant coverage of a
planar interface, or coverage per substrate area \cite{ThAP2012pf}. Similarly, one
must replace $\phi \to \psi/h$ before the variation of $\mathcal{F}_\mathrm{b}$
is computed \cite{Thie2011epjst}. This yields
\begin{eqnarray}
p &=& \frac{\delta \mathcal{F}}{\delta h} =-\partial_\alpha \left( \sqrt{a} \,\sigma^{\alpha\beta}\partial_\beta h \right)  - \Pi(h) + p_\mathrm{b},
 \mylab{eq:insolvarh}\\
 p_\mathrm{b} &=& \frac{\delta \mathcal{F}_\mathrm{b}}{\delta h} = p_\mathrm{osm}
 {+} \kappa\left[\frac {1}{2} |\nabla \phi|^2 +\frac {\phi}{h}
                  \nabla\cdot( h \nabla \phi) \right],  \quad  p_\mathrm{osm} = f(\phi) -\phi f'(\phi).
\mylab{eq:solvarh}
\end{eqnarray}
where $p_\mathrm{osm}$ is the bulk osmotic pressure. With solely the ideal-gas
(entropic) terms in Eq.~(\ref{eq:enfbulk0}), it becomes
$p_\mathrm{osm}= -kT \phi/l^3$. Note too that $\nabla$ is the 2D gradient
operator, and $\nabla^2$ is the 2D Laplacian.  The second term in Eq.~(\ref{eq:insolvarh})
is the disjoining pressure, while the first term contains the
interfacial stress
\begin{eqnarray}
\sigma^{\alpha\beta} &=& \frac{\delta \mathcal{F}_\mathrm{s}}{\delta a_{\alpha\beta}}=
\frac{1}{2}a^{\alpha\beta} [ f_\mathrm{s} (\Gamma) - \Gamma f_\mathrm{s}' (\Gamma)] -
\frac{\kappa_\mathrm{s}}{4} a^{\alpha\beta} a^{\gamma\delta}\partial_\gamma\Gamma \partial_\delta\Gamma  \nonumber \\ 
 &+& \frac{\kappa_\mathrm{s} \Gamma}{4} a^{\alpha\beta} \left[
\partial_\gamma \left( \sqrt{a}\, a^{\gamma\delta}\partial_\delta\Gamma\right) + 
\partial_\delta\left(\sqrt{a}\,  a^{\gamma\delta}\partial_\gamma\Gamma\right)\right].
\mylab{eq:insolvar}
\end{eqnarray}
In particular, the standard surface tension is defined as
\begin{equation}
\gamma(\Gamma) = a_{\alpha\beta}(\sigma^{\alpha\beta})_{\kappa_\mathrm{s} \to
  0}= f_\mathrm{s} (\Gamma) - \Gamma f_\mathrm{s}'(\Gamma).
\mylab{eq:surftens-general}
\end{equation}

The function $f_\mathrm{s}$ in Eq.~(\ref{eq:enfsurf0}) with
Eq.~(\ref{eq:enfsurf}) for $b=0$ results then in what is sometimes called the
Langmuir equation of state \cite{PaSt1996pf,KrVM2004pf} or the Von
Szyckowski equation \cite{EgPS1999jfm}
\begin{equation}
\gamma=\gamma_0+\frac{kT}{l_\mathrm{s}^2} \ln(1-\Gamma),
\mylab{eq:surftens-langmuir}
\end{equation}
i.e., for
$\Gamma\ll1$ one has $\gamma\approx\gamma_0-kT\Gamma/l_\mathrm{s}^2=\gamma_0-\gamma_\Gamma\Gamma$, where we introduced the Marangoni coefficient $\gamma_\Gamma=kT/l_\mathrm{s}^2$ 
for the resulting linear solutal Marangoni effect.
Note that with $b\neq 0$ in
Eq.~(\ref{eq:enfsurf}) one obtains the Frumkin equation of state as given in
\cite{PaSt1996pf} and further discussed below in section~\ref{sec:surfsol-nonconsfluxadsorp}.

The surface chemical potential $\mu_\mathrm{s}$ is obtained by varying Eq.~(\ref{eq:ff})
with respect to $\widetilde{\Gamma}$:
\begin{equation}
\mu_\mathrm{s} = \frac{\delta \mathcal{F}}{\delta\widetilde{\Gamma}} = \frac{df_\mathrm{s}}{d\Gamma} -  \frac{\kappa_\mathrm{s}}{2}\left[
\partial_\alpha\left( a^{\alpha\beta} \partial_\beta\Gamma\right) +\partial_\beta\left( a^{\alpha\beta} \partial_\alpha\Gamma\right) \right].
\mylab{eq:musov}
\end{equation}
Finally, the  bulk chemical potential is \cite{ThAP2016note1}
\begin{equation}
\mu = \frac{\delta \mathcal{F}}{\delta \psi}= f'(\phi) 
- \kappa h^{-1}  \nabla \cdot  \left(h\nabla\phi \right).
\mylab{eq:mu}
\end{equation}

The mechanical interaction between the surfactant layer and the bulk liquid is carried by the balance of the interfacial stress and the viscous stress in the bulk fluid proportional to the normal derivative of the  velocity $v^\alpha$ tangential to the interface and the bulk viscosity $\eta$:
\begin{equation}
\sigma^{\alpha\beta}{}_{;\beta}  = \eta v^\alpha{}_{;n},
\mylab{eq:tang}
\end{equation}
where the semicolon denotes the covariant derivative necessary when vectors defined on a curved interface are involved. This equation reduces to the commonly used tangential stress balance including the Marangoni force when the rigidity $\kappa_\mathrm{s}$ is neglected. 

\subsection{Thin film hydrodynamics}
\mylab{sec:surfsol:tfh}
The above general expressions for the surface stress and pressure can
be simplified in the case when the curvature and inclination are small so
that the long-wave or lubrication approximation can be made. To this end,
we scale $\partial_\alpha \sim O(\epsilon)$, $v^\alpha 
\sim O(\epsilon)$, $\partial_t \sim O(\epsilon^2)$ and retain terms up
to the lowest relevant order in $\epsilon \ll 1$. With this scaling,
$a_{\alpha\beta}$ differs from the Cartesian surface metric
$\delta_{\alpha\beta}$ by $O(\epsilon^2)$, so that
$a_{\alpha\beta}=\delta_{\alpha\beta}+\epsilon^2\partial_\alpha
h \partial_\beta h$ and its inverse is, to leading order,
$a^{\alpha\beta}=\delta_{\alpha\beta}-\epsilon^2\partial_\alpha
h \partial_\beta h$.  Then, the above expressions can be rewritten
using Cartesian coordinates $x_\alpha$ spanning the plane of the
substrate, whereby the distinction between covariant and contravariant
tensors disappears (so that all indices can be written as subscripts)
and covariant derivatives are replaced by usual partial
derivatives. Retaining the leading order terms only,
Eqs.~(\ref{eq:insolvarh}) -- (\ref{eq:musov}) become 
\begin{eqnarray}
 && p = \frac{\delta \mathcal{F}}{\delta h} = -\nabla \cdot \left[ (\gamma_0-p_\mathrm{s}) \nabla h \right]
- \Pi(h)  + p_\mathrm{b},
\mylab{eq:insolvarh0}\\
&&p_\mathrm{s} = \gamma_0 - \gamma(\Gamma) 
- \kappa_\mathrm{s} \left(\Gamma \nabla^2 \Gamma -\frac{1}{2} |\nabla\Gamma |^2 \right) \mylab{eq:insolvar0}\\
&&\mu_\mathrm{s} = \frac{\delta \mathcal{F}}{\delta\widetilde{\Gamma}}= f_\mathrm{s}'(\Gamma) -  \kappa_\mathrm{s} \nabla^2 \Gamma,
\mylab{eq:mus0}
\end{eqnarray}
where we have used $\sigma_{\alpha\beta}  = \delta_{\alpha\beta}\, (\gamma_0-p_\mathrm{s})$ and where $p_\mathrm{s}$ is the surface pressure that captures the
difference between reference surface tension without surfactant
$\gamma_0$ and the full concentration-dependent expression (including rigidity).
Further, $p_\mathrm{b}$ and  $p_\mathrm{osm}$
remain as in Eq.~(\ref{eq:solvarh}) while $\mu$ is still given by Eq.~(\ref{eq:mu}).

The bulk flow field is computed by solving the modified Stokes
equation, also called the momentum equation of model-H
\cite{JaVi1996pf,ThMF2007pf}. Its relevant components are parallel to the
substrate plane:
\begin{equation}
 \eta \mathbf{v}''(z)= \nabla {p} + \phi \nabla \mu.
 \mylab{eq:stokes}
\end{equation}
where $\mathbf{v}$ is the 2D vector of the velocities parallel to the
substrate plane.
An alternative form of equation~(\ref{eq:stokes}) can be obtained using the relation
\begin{equation}
\nabla p_\mathrm{b} = f''(\phi) \nabla \phi -\kappa \nabla\left[\frac {1}{2} |\nabla \phi|^2 +\frac {\phi}{h} \nabla\cdot( h \nabla \phi) \right]
=-\phi\nabla{\mu} + \frac{\kappa}{h} \nabla(h |\nabla \phi|^2),
\mylab{eq:nabs}
\end{equation}
which reduces the right-hand side of the Stokes equation~(\ref{eq:stokes}) to 
\begin{equation}
\nabla(p-p_\mathrm{b})+\nabla p_\mathrm{b} + \phi \nabla \mu = \nabla\widehat{p} + \frac{\kappa}{h} \nabla(h |\nabla \phi|^2),
\mylab{eq:nabs1}
\end{equation}
where $\widehat{p}$ is the effective pressure excluding $p_\mathrm{b}$. This
shows that osmotic pressure $p_\mathrm{osm}$ does not affect
hydrodynamic flow (as $\nabla p_\mathrm{osm}=-\phi\nabla\mu$), while
the contribution of the bulk rigidity is expressed by the last term in the above relation. 

Solving Eq.~(\ref{eq:stokes}) in the lubrication approximation with the
no-slip 
boundary condition at the substrate plane $z=0$ and the momentum
balance condition (\ref{eq:tang}) at the interface $z=h$ yields
\begin{equation}
 \mathbf{v} =-\frac{z}{\eta}\left[\nabla p_\mathrm{s} +
 \left(h - \frac z2 \right) (\nabla {p} + \phi \nabla \mu) \right] .
 \mylab{eq:vel}
\end{equation}
Integrated over the local film thickness, this leads to the convective fluid flux
\begin{equation}
\mathbf{J}_\mathrm{conv} = \int_0^h \mathbf{v} dz =
 -\frac{h^2}{2\eta}\nabla p_\mathrm{s} -
 \frac{h^3}{3\eta} (\nabla {p} + \phi \nabla \mu) ,
 %+ \partial_\beta T_{\alpha\beta} \right],
 \mylab{eq:vel}
\end{equation}
and the interfacial velocity $\mathbf{v}_\mathrm{s} = \mathbf{v}(h)$. 
Then the volume conservation condition
\begin{equation}
\partial_t h=-\nabla\cdot\mathbf{J}_\mathrm{conv} 
 \mylab{eq:htcons}
\end{equation}
results, to leading order, in the evolution equation of the film thickness
\begin{equation}
\partial_t h = \nabla \cdot
\left[ \frac {h^3}{3 \eta}(\nabla {p} + \phi \nabla \mu) 
 + \frac{h^2}{2 \eta}\nabla p_\mathrm{s} \right] 
 - J_\mathrm{ev}(h,\Gamma,\phi),
\mylab{eq:lub}
\end{equation}
where we now incorporated the evaporation flux $J_\mathrm{ev}$. The leading-order equations expressing the surface and bulk surfactant conservation laws are  
\begin{eqnarray}
\partial_t \Gamma &=&  \nabla \cdot
\left( \frac {h^2}{2\eta}\Gamma(\nabla {p} + \phi \nabla \mu)
 +  \frac {h }{\eta}\Gamma \nabla p_\mathrm{s} + M_\mathrm{s}(\Gamma) \nabla \mu_\mathrm{s} \right) 
 + \widetilde{J}_\mathrm{ad}(\Gamma,\phi),
\mylab{eq:gameq}\\
\partial_t \psi &=&  \nabla  \cdot
\left( \frac {h^2}{3\eta}\psi(\nabla {p} + \phi \nabla \mu) 
 + \frac{h}{2\eta} \psi \nabla p_\mathrm{s}+ h M(\phi) \nabla \mu \right) 
 - J_\mathrm{ad}(\Gamma,\phi),
\mylab{eq:psieq}
\end{eqnarray}
where $M_\mathrm{s}(\Gamma)$ and $M(\phi)$ are general surface and
bulk mobility functions and
$\tilde J_\mathrm{ad}= J_\mathrm{ad}/l_\mathrm{s}$ is the net
surfactant adsorption flux; surface distortions contribute to
Eq.~(\ref{eq:gameq}) as $O(\epsilon^2)$ terms only.  In the dilute
limit, the mobilities can be expressed as
\begin{equation}
M_\mathrm{s}(\Gamma)=\frac{D_\mathrm{s} l_\mathrm{s}^2\Gamma}{kT} \quad\mathrm{and}\quad
M(\phi)= \frac{D l^3 \phi}{kT} 
\mylab{eq:molmobs}
\end{equation}
where ${D}_\mathrm{s}$ and ${D}$  are surface
and bulk diffusivities, respectively. The lengths in the diffusion
terms are introduced for convenience. They ensure that the
diffusivities $D$'s have units m$^2/$s as usual for diffusion
constants.
The conserved dynamics in Eqs.~(\ref{eq:gameq}) and (\ref{eq:psieq})
have the form of conservation laws 
\begin{equation}
\partial_t\Gamma = -\nabla\cdot(\Gamma\vec{v}_\mathrm{s} +
\vec{J}^\Gamma_\mathrm{diff}) \qquad\mbox{and}
\mylab{eq:gammatcons}
\end{equation}
\begin{equation}
\partial_t(\phi h) =-\nabla\cdot(\phi\vec{J}_\mathrm{conv} +
\vec{J}^\phi_\mathrm{diff}), 
\mylab{eq:psitcons}
\end{equation}
respectively.
We also take into account the relation
\begin{equation}
\nabla p_\mathrm{s} = \Gamma f_\mathrm{s}''(\Gamma)\nabla \Gamma + \kappa_\mathrm{s}\nabla \left(\Gamma \nabla^2 \Gamma -\frac{1}{2} |\nabla\Gamma |^2 \right) 
= \Gamma \nabla  f_\mathrm{s}' (\Gamma)
-\kappa_\mathrm{s}\Gamma \nabla \nabla^2 \Gamma =\Gamma\nabla \mu_\mathrm{s},
\mylab{eq:nab} \end{equation}
that allows us to replace the gradient of the surface pressure in Eq.~(\ref{eq:lub}) by $\nabla \mu_\mathrm{s}$
\subsection{Gradient dynamics formulation}
\mylab{sec:surfsol:graddyn}
Eqs.~(\ref{eq:gameq}) -- (\ref{eq:psieq}) can be now presented in
the general gradient dynamics form~(\ref{eq:nn3}) with $a, b = 1,2,3$ for three fields as
\begin{eqnarray}
\partial_t h &=&  \nabla \cdot
\left( Q_{hh}\nabla \frac{\delta \mathcal{F}}{\delta h}  
 + Q_{h\Gamma}\nabla \frac{\delta \mathcal{F}}{\delta \widetilde\Gamma} 
 + Q_{h\psi} \nabla \frac{\delta \mathcal{F}}{\delta \psi} \right) 
 - \beta_\mathrm{evap}\left(  \frac{\delta \mathcal{F}}{\delta h} - p_\mathrm{vap} \right),
\mylab{eq:lubons} \\
\partial_t \Gamma &=&  \nabla \cdot
\left( Q_{\Gamma h}\nabla \frac{\delta \mathcal{F}}{\delta h} 
  + Q_{\Gamma\Gamma}\nabla \frac{\delta \mathcal{F}}{\delta \widetilde\Gamma} 
  + Q_{\Gamma\psi} \nabla \frac{\delta \mathcal{F}}{\delta \psi}\right) 
- \beta_\mathrm{\psi\Gamma}\left(  \frac{1}{l_\mathrm{s}}\frac{\delta
                      \mathcal{F}}{\delta \widetilde\Gamma} - \frac{\delta \mathcal{F}}{\delta \psi} \right),
\mylab{eq:gamons}\\
\partial_t \psi &=&  \nabla  \cdot
\left( Q_{\psi h}\nabla \frac{\delta \mathcal{F}}{\delta h} 
  + Q_{\psi\Gamma}\nabla \frac{\delta \mathcal{F}}{\delta \widetilde\Gamma}+ Q_{\psi\psi} \nabla\frac{\delta \mathcal{F}}{\delta \psi}\right) 
- \beta_\mathrm{\psi\Gamma}\left(  l_\mathrm{s} \frac{\delta \mathcal{F}}{\delta \psi} - \frac{\delta \mathcal{F}}{\delta \widetilde\Gamma} \right),
\mylab{eq:psions}
\end{eqnarray}
The mobility matrix for the conserved dynamics reads
\begin{equation}
\mathbf{Q}^\mathrm{c}\,=\,\left( 
\begin{array}{ccc}  
Q^\mathrm{c}_{hh} & Q^\mathrm{c}_{h \Gamma} & Q^\mathrm{c}_{h \psi} \\[.3ex]
Q^\mathrm{c}_{\Gamma h}  &Q^\mathrm{c}_{\Gamma \Gamma} & Q^\mathrm{c}_{\Gamma \psi }\\[.3ex]
Q^\mathrm{c}_{\psi h} & Q^\mathrm{c}_{\psi \Gamma} & Q^\mathrm{c}_{\psi \psi} 
\end{array}
\right)
\,=\,\left( 
\begin{array}{ccc}  
\frac{h^3}{3 \eta} & \frac{h^2\Gamma}{2 \eta} & \frac{h^2\psi}{3 \eta} \\[.3ex]
\frac{h^2\Gamma}{2 \eta} &\frac{h\Gamma^2}{\eta}+ M_\mathrm{s}(\Gamma) & \frac{h\psi\Gamma}{2 \eta}\\[.3ex]
\frac{h^2\psi}{3 \eta} & \frac{h\psi\Gamma}{2 \eta}  &
                                                       \frac{h\psi^2}{3 \eta}+ h M(\phi).
\end{array}
\right)
\mylab{eq:solsurf-mob}
\end{equation}
Note that $\mathbf{Q}^\mathrm{c}$ is symmetric and positive definite,
corresponding to Onsager relations between the fluxes and positive
entropy production, respectively. Also the mobility matrix 
\begin{equation}
\mathbf{Q}^\mathrm{nc}\,=\,\left( 
\begin{array}{ccc}  
Q^\mathrm{nc}_{hh} & Q^\mathrm{nc}_{h \Gamma} & Q^\mathrm{nc}_{h \psi} \\[.3ex]
Q^\mathrm{nc}_{\Gamma h}  &Q^\mathrm{nc}_{\Gamma \Gamma} & Q^\mathrm{nc}_{\Gamma \psi }\\[.3ex]
Q^\mathrm{nc}_{\psi h} & Q^\mathrm{nc}_{\psi \Gamma} & Q^\mathrm{nc}_{\psi \psi} 
\end{array}
\right)
\,=\,\left( 
\begin{array}{ccc}  
\beta_\mathrm{evap}& 0 & 0 \\[.3ex]
0&\frac{\beta_\mathrm{\psi\Gamma}}{l_\mathrm{s}} & -\beta_\mathrm{\psi\Gamma}\\[.3ex]
0 & -\beta_\mathrm{\psi\Gamma} & l_\mathrm{s}\beta_\mathrm{\psi\Gamma}.
\end{array}
\right)
\mylab{eq:solsurf-mob-nc}
\end{equation}
for the non-conserved dynamics is symmetric and positive definite.
Note that the mobility functions that involve $\Gamma$ have a
dimension different from the other terms; the same applies to the 
variations. However, the overall contributions to the respective fluxes 
of course have the same dimensions.

The final non-conserved terms in Eqs.~(\ref{eq:lubons}) to
(\ref{eq:psions}) correspond to $-J_\mathrm{evap}$, $\widetilde
J_\mathrm{ad}=J_\mathrm{ad}/l_\mathrm{s}$, and 
$-J_\mathrm{ad}$, respectively. We discuss below in
Section~\ref{sec:surfsol-nonconsfluxadsorp} that in the limit of a flat
surface and without rigidity terms they give exactly the expressions
for adsorption/desorption most often derived in the literature from kinetic
considerations \cite{Leal2007,AtkinsPaula2010}. However, in contrast to these
considerations, our formulation also naturally captures the influence
of surface modulations and rigidity effects.

Comparing the three conserved fluxes in
Eqs.~(\ref{eq:lubons})-(\ref{eq:psions}) to the conservation laws
Eqs.~(\ref{eq:htcons}),~(\ref{eq:gammatcons}), and~(\ref{eq:psitcons})
one notes that only $Q^\mathrm{c}_1=Q^\mathrm{c}_{hh}$,
$Q^\mathrm{c}_2=Q^\mathrm{c}_{h\Gamma}$ and
$Q^\mathrm{c}_3=Q^\mathrm{c}_{\Gamma \Gamma}$ are independent, the
other mobility functions can be derived from the relation between
$\vec{J}_\mathrm{conv}$ and $\phi\vec{J}_\mathrm{conv}$, i.e., the
mobility matrix is
\begin{equation}
\mathbf{Q}^\mathrm{c}\,=\,\left( 
\begin{array}{ccc}  
Q^\mathrm{c}_1 &  Q^\mathrm{c}_2 & \phi Q^\mathrm{c}_1\\[.3ex]
Q^\mathrm{c}_2 & Q^\mathrm{c}_3 &\phi Q^\mathrm{c}_2\\[.3ex]
\phi Q^\mathrm{c}_1 &\phi Q^\mathrm{c}_2 & \phi^2 Q^\mathrm{c}_1
\end{array}
\right)
\mylab{eq:solsurf-mob2}
\end{equation}
This structure ensures that for any $f(\phi)$ the osmotic pressure in
the bulk film $p_\mathrm{osm}$ does not contribute to the convective
flux $J_\mathrm{conv}$. However, it does have an influence on evaporation
(see section~\ref{sec:surfsol-nonconsflux1}).
Without slip, one has $Q^\mathrm{c}_1=h^3/3\eta$,
$Q^\mathrm{c}_2=\Gamma h^2/2\eta$ and $Q^\mathrm{c}_3=\Gamma^2 h/\eta$, but slip can be
easily incorporated.
%
%%%%%%%%%%%%%%%%%%%%%%%%%%%%%%%%%%%%%%%%%%%%%%%%%%%%%%%%%%%%%%%%%%%%%%%%%%%%%%%
\subsection{Non-conserved fluxes}
\mylab{sec:surfsol-nonconsflux1}
%%%%%%%%%%%%%%%%%%%%%%%%%%%%%%%%%%%%%%%%%%%%%%%%%%%%%%%%%%%%%%%%%%%%%%%%%%%%%%%
%
The general gradient dynamics form in
Eqs.~(\ref{eq:lubons})-(\ref{eq:psions}) incorporates conserved and
non-conserved fluxes. The considered non-conserved fluxes include an
evaporation/condensation flux $ J_\mathrm{ev}$ that only enters the
equation for the film height (\ref{eq:lubons}) and an
adsorption/desorption flux $ J_\mathrm{ad}$ that enters the equations
for the bulk and surface concentrations (\ref{eq:gamons}) and
(\ref{eq:psions}). If the respective fluxes are zero the exchange
processes are at equilibrium, i.e., the evaporation and
condensation of the solvent balance as well as adsorption and
desorption of the solute. In the following we discuss the fluxes
individually.
%
%%%%%%%%%%%%%%%%%%%%%%%%%%%%%%%%%%%%%%%%%%%%%%%%%%%%%%%%%%%%%%%%%%%%%%%%%%%%%%%
\subsubsection{Evaporation and condensation}
\mylab{sec:surfsol-nonconsfluxevap}
%%%%%%%%%%%%%%%%%%%%%%%%%%%%%%%%%%%%%%%%%%%%%%%%%%%%%%%%%%%%%%%%%%%%%%%%%%%%%%%
%
Assuming that the solute does not influence the
film height the evaporation flux is given by
\begin{equation}
 J_\mathrm{ev}(h,\Gamma,\phi)=\beta_\mathrm{evap}\left(  \frac{\delta \mathcal{F}}{\delta h} -p_\mathrm{vap} \right).
\mylab{eq:evapflux} 
\end{equation}
With (\ref{eq:insolvarh0}) and (\ref{eq:solvarh}) this becomes 
\begin{equation}
 J_\mathrm{ev}(h,\Gamma,\phi)=\beta_\mathrm{evap}\left(  \nabla \cdot
   \left(p_\mathrm{s} \nabla h \right)  - \Pi(h)  + p_\mathrm{osm}
 -\kappa\left[\frac {1}{2} |\nabla \phi|^2 +\frac {\phi}{h} \nabla\cdot( h \nabla \phi) \right] -p_\mathrm{vap} \right),
\mylab{eq:evapflux} 
\end{equation}
where as before $p_\mathrm{osm} = f(\phi) -\phi f'(\phi)$ and
$p_\mathrm{vap}$ is the partial vapour pressure in the ambient air. Besides the
known Kelvin effect (first term on the r.h.s., here with the full dependence
$p_\mathrm{s}(\Gamma)$) \cite{ReCo2013pre}, wettability (second term on the
r.h.s.) and osmotic pressure (third term) influence evaporation as
does the bulk rigidity (fourth term). Normally, even on
mesoscopic scales, the dominant term is that involving
the vapour pressure (fifth term) and this term largely controls the
evaporation rate -- see Ref.~\cite{Thie2014acis} for further discussion on this.
However, the other terms do matter close to contact
lines, for nanodroplets and at diffuse interfaces of dense and dilute
phases. Note that such thermodynamically consistent relations for
$J_\mathrm{ev}$ are also obtained for all the model extensions discussed below
in section~\ref{sec:surfsol-special}. Also note that the rate
$\beta_\mathrm{evap}$ is not necessarily constant. It may depend on
film height, e.g., $\beta_\mathrm{evap}=E/(K+h)$ when incorporating 
effects of latent heat \cite{AjHo2001jcis,Ajae2005pre,ReCo2010mst} (see
\cite{Thie2014acis} for more details).

Problems may arise in the limit of very high bulk concentrations of
the solute, since the physical film height can then be virtually
identical to the effective solute height in contradiction to the model
assumption that the effective solute height is small as compared to
the effective solvent height that is identified with the film height.
This issue may be resolved through a solvent-solute symmetric model as
proposed in \cite{XuTQ2015jpcm} in the two-field case. This case of
high solute concentrations will be pursued elsewhere.
%
%%%%%%%%%%%%%%%%%%%%%%%%%%%%%%%%%%%%%%%%%%%%%%%%%%%%%%%%%%%%%%%%%%%%%%%%%%%%%%%
\subsubsection{Adsorption and desorption}
\mylab{sec:surfsol-nonconsfluxadsorp}
%%%%%%%%%%%%%%%%%%%%%%%%%%%%%%%%%%%%%%%%%%%%%%%%%%%%%%%%%%%%%%%%%%%%%%%%%%%%%%%
%
Besides evaporation, the non-conserved part of the gradient dynamics
(\ref{eq:lubons})-(\ref{eq:psions}) also describes the dynamics of
exchange of surfactant molecules between the liquid bulk and the free surface. When
$\widetilde J_\mathrm{ad}>0$ this corresponds to an adsorption flux of molecules
attaching to the free surface, while when $\widetilde J_\mathrm{ad}<0$
there is desorption from the free surface, i.e., it is an influx
into the bulk. Overall, the exchange between the bulk and the free surface is
mass conserving, i.e., it suffices to discuss $\widetilde
J_\mathrm{ad}$, then $J_\mathrm{ad}=l_\mathrm{s}\widetilde
J_\mathrm{ad}$. Within the gradient dynamics it is given by
\begin{eqnarray}
\widetilde J_\mathrm{ad}(h,\Gamma,\phi)&=&
 \beta_\mathrm{\psi\Gamma}\left(  \frac{\delta \mathcal{F}}{\delta \psi} - \frac{1}{l_\mathrm{s}}\frac{\delta \mathcal{F}}{\delta \widetilde\Gamma} \right)
\mylab{eq:exflux1} \\
&=&\beta_\mathrm{\psi\Gamma}\left( \mu - \frac{1}{l_\mathrm{s}}\mu_\mathrm{s}\right)\mylab{eq:exflux2} \\
&=& \beta_\mathrm{\psi\Gamma}\left[  \frac{df}{d\phi} - \kappa h^{-1}
  \nabla \cdot  \left(h\nabla\phi \right) -\frac{1}{l_\mathrm{s}}
\left( \frac{df_\mathrm{s}}{d\Gamma} -  \kappa_\mathrm{s} \nabla^2 \Gamma\right)\right],
\mylab{eq:exflux3}
\end{eqnarray}
where we have used Eqs.~(\ref{eq:mu}) and ~(\ref{eq:mus0}). Note that
the bulk rigidity  ($\kappa\neq0$) introduces an explicit film height
dependence. Without rigidity influences ($\kappa, \kappa_\mathrm{s}=0$), the flux is 
$\widetilde J_\mathrm{ad}=\beta_\mathrm{\psi\Gamma}\left[  \frac{df}{d\phi} -\frac{1}{l_\mathrm{s}}
\frac{df_\mathrm{s}}{d\Gamma}\right]$, and one may now consider several particular cases.

In the dilute limit for the bulk concentration
$\phi$ we have $f_b^{ex}=0$ and Eq.~(\ref{eq:enfbulk0}) becomes
\begin{equation}
f(\phi)=\frac{kT}{l^{3}}[\phi(\ln\phi-1)].
\mylab{eq:bulk-dilute}
\end{equation}
This implies that when solely entropic surface packing effects are included
in $f_\mathrm{s}(\Gamma)$, i.e., Eqs.~(\ref{eq:enfsurf0}) and
(\ref{eq:enfsurf}) with inter-molecular attraction parameter $b=0$, we obtain
\begin{equation}
\widetilde J_\mathrm{ad} = \beta_\mathrm{\psi\Gamma}
  \frac{kT}{l_\mathrm{s}^{3}}\ln\frac{(1-\Gamma)\phi}{\Gamma},
\mylab{eq:adsorpt-flux2}
\end{equation}
where we also assume $l=l_\mathrm{s}$ (otherwise
$\phi\to\phi^{l_\mathrm{s}^3/l^3}$).  An expression identical to
(\ref{eq:adsorpt-flux2}) is given in section~2.3 of \cite{DiAn1996jpc}
where a free energy approach is followed to study the kinetics of
surfactant adsorption (set $\beta=0$ in Eq.~(2.14) to recover the purely
entropic case). For a full agreement with \cite{DiAn1996jpc} one needs
$\beta_\mathrm{\psi\Gamma}=\widetilde M\phi$ where $\widetilde M$ is a
molecular mobility. The approximation discussed next makes it likely
that there is actually a typo in \cite{DiAn1996jpc} and it should read
$\beta_\mathrm{\psi\Gamma}=\widetilde M\Gamma$. 

In many cases, the surfactant isotherms that relate equilibrium
surface concentration $\Gamma_\mathrm{eq}$ and equilibrium bulk
concentration $\phi_\mathrm{eq}$ are introduced based on kinetic
arguments of equal desorption and adsorption fluxes (see, e.g.,
Refs.~\cite{Leal2007,AtkinsPaula2010}). However, the isotherm is an
equilibrium property and may be directly obtained from the free
energy. In the present context, one has at equilibrium $\widetilde J_\mathrm{ad}
=0$, i.e.,
$\phi_\mathrm{eq}=\Gamma_\mathrm{eq}/(1-\Gamma_\mathrm{eq})$ or
$\Gamma_\mathrm{eq}=\phi_\mathrm{eq}/(1+\phi_\mathrm{eq})$
corresponding to the Langmuir adsorption isotherm \cite{Leal2007}. To obtain
the kinetics when the system is out-of but still close to equilibrium we
expand the logarithm in Eq.~(\ref{eq:adsorpt-flux2}) about the equilibrium state and obtain
\begin{equation}
\widetilde J_\mathrm{ad} \approx \beta_\mathrm{\psi\Gamma}
  \frac{kT}{\Gamma l_\mathrm{s}^{3}} \left[(1-\Gamma)\phi - \Gamma\right].
\mylab{eq:adsorpt-flux3}
\end{equation}
This expression for the effective adsorption flux (adsorption minus
desorption) agrees for $\beta_\mathrm{\psi\Gamma}=\widetilde M\Gamma$
up to normalisation factors with Eqs.~(6) of Ref.~\cite{PaSt1996pf}
that result from kinetic considerations.  

One may also go beyond purely entropic interactions, e.g., by using
Eq.~(\ref{eq:enfsurf}) or other forms of $f_\mathrm{s}^{ex}(\Gamma)$.
With $b>0$ in Eq.~(\ref{eq:enfsurf}) one introduces a simple attraction between surfactant
molecules at the free surface. Then 
\begin{equation}
\widetilde J_\mathrm{ad} = \beta_\mathrm{\psi\Gamma}
  \frac{kT}{l_\mathrm{s}^{3}}\ln\frac{(1-\Gamma)\phi^{l_\mathrm{s}^3/l^3}}{\Gamma} + \beta_\mathrm{\psi\Gamma}\frac{b\Gamma}{l_\mathrm{s}},
\mylab{eq:adsorpt-flux2frumkin}
\end{equation}
where this time we retain the general $l\neq l_\mathrm{s}$.

At equilibrium $\widetilde J_\mathrm{ad}=0$, i.e.,
\begin{equation}
\phi_\mathrm{eq}=\left(\frac{\Gamma_\mathrm{eq}}{1-\Gamma_\mathrm{eq}}\right)^{(l/l_\mathrm{s})^3}\,e^{
  -\tilde b\Gamma_\mathrm{eq}}
\mylab{eq:phi-gamma-frumkin}
\end{equation}
where $\tilde b= bl^3/kTl_\mathrm{s}$, or in an implicit form
\begin{equation}
\Gamma_\mathrm{eq}=\frac{\phi_\mathrm{eq}^{(l_\mathrm{s}/l)^3} e^{\tilde
    b (l_\mathrm{s}/l)^3\Gamma_\mathrm{eq}}}{1+\phi_\mathrm{eq}^{(l_\mathrm{s}/l)^3}  e^{\tilde b (l_\mathrm{s}/l)^3\Gamma_\mathrm{eq}}}.
\mylab{eq:phi-gamma2-frumkin}
\end{equation}
Both are common in the literature \cite{FZLM2003jcis}, in particular, for
$l=l_\mathrm{s}$ they are known as the Frumkin isotherm \cite[chap.~2.N]{Leal2007}:
\begin{equation}
\phi_\mathrm{eq}=\left(\frac{\Gamma_\mathrm{eq}}{1-\Gamma_\mathrm{eq}}\right)\,e^{-\tilde
  b\Gamma_\mathrm{eq}}
\mylab{eq:phi-gamma-b}
\end{equation}
or
\begin{equation}
\Gamma_\mathrm{eq}=\frac{\phi_\mathrm{eq}}{e^{-\tilde b \Gamma_\mathrm{eq}}+\phi_\mathrm{eq} }.
\mylab{eq:phi-gamma2-b}
\end{equation}
The kinetic adsorption equation given in \cite{FeSt1999jcis} is obtained by expanding (\ref{eq:adsorpt-flux2frumkin})
about this equilibrium state (\ref{eq:phi-gamma-b}).  The linearised
flux is
\begin{equation}
\widetilde J_\mathrm{ad} = \beta_\mathrm{\psi\Gamma}
  \frac{kT}{l^{3}\Gamma} e^{\tilde b\Gamma} \left[(1-\Gamma)\phi -\Gamma e^{-\tilde b\Gamma}\right]
\mylab{eq:adsorpt-flux2frumkin-lin}
\end{equation}
that has the same form as Eq.~(16) of Ref.~\cite{FeSt1999jcis} and
implies certain $\Gamma$-dependencies of their mobilities $\alpha$ and
$\beta$ or of our mobility $\beta_\mathrm{\psi\Gamma}$. Note that the case of adhesion (their $K<0$) here corresponds
to $\tilde b>0$.

The expression in Eq.~\eqref{eq:evapflux}, that is linear in the
thermodynamic potentials (variations of $\mathcal{F}$) must
be linearised about the equilibrium state $\widetilde J_\mathrm{ad}=0$ to obtain
  the expressions obtained in the literature based on kinetic
  considerations. This may imply that these kinetic considerations
  only capture a linearised picture of the process.  Alternatively,
  one may introduce expressions such as
  $(\phi-\Gamma)/(\log\phi-\log\Gamma)$ into the mobility
  $\beta_\mathrm{\psi\Gamma}$ as proposed in \cite{Miel2011n} in the
  context of gradient dynamics formulations of reaction-diffusion
  dynamics. However, for the more complicated free energies discussed
  here this seems inadequate.  Another option is to go beyond linear
  nonequilibrium thermodynamics, i.e., beyond the expression linear in
  the thermodynamic potentials in
  Eq.~(\ref{eq:evapflux}). For activated processes, activation
  barriers have to be overcome and Arrhenius-type exponential factors
  may be appropriate. For instance, an adsorption flux
\begin{equation}
 \widetilde J_\mathrm{ad}(h,\Gamma,\phi)=
 \hat\beta_\mathrm{\psi\Gamma}\left(\exp \left[  -\frac{a^3}{kT}\frac{\delta \mathcal{F}}{\delta \psi}
+\frac{a^3}{kTl_\mathrm{s}}\frac{\delta \mathcal{F}}{\delta \widetilde\Gamma} \right] -1 \right)
\mylab{eq:exflux1arrhenius2} 
\end{equation}
($a$ is a microscopic length scale) with appropriately
defined mobility $\hat\beta_\mathrm{\psi\Gamma}$ results in the 
same expressions for the flux as obtained via kinetic considerations.

We end this section with a side remark on the general adsorption isotherm.
Using the standard definition of the surface tension given in Eq.~(\ref{eq:surftens-general}), we obtain
\begin{equation}
d\gamma = -\Gamma_\mathrm{eq} f''_\mathrm{s}\,d\Gamma_\mathrm{eq} = -\Gamma_\mathrm{eq} f''_\mathrm{s}\,
\frac{d\Gamma_\mathrm{eq}}{d (\ln\phi_\mathrm{eq})} d (\ln\phi_\mathrm{eq})
\mylab{eq:gibbs-adsorp0}
\end{equation}
In the dilute limit of the bulk surfactant concentration,
i.e.\ for$f(\phi) =\frac{kT}{l^{3}}\phi(\ln\phi-1)$,
the adsorption isotherm is $(kTl_\mathrm{s}/l^3)\ln\phi_\mathrm{eq}=f'_\mathrm{s}$,
i.e., $d(\ln\phi_\mathrm{eq})/d\Gamma_\mathrm{eq} = (l^3/kTl_\mathrm{s}) f''_\mathrm{s}$ implying that the Gibbs
adsorption isotherm
\begin{equation}
d\gamma = - \frac{kT l_\mathrm{s}}{l^3}\Gamma_\mathrm{eq} \,d(\ln\phi_\mathrm{eq})
\mylab{eq:gibbs-adsorp}
\end{equation}
is valid for any form of $f''_\mathrm{s}(\Gamma)$. However, this is
not the case for more complicated expressions for $f(\phi)$ or, indeed, when rigidity
effects are included. Then Eq.~(\ref{eq:exflux3}) with
$\widetilde J_\mathrm{ad}=0$ provides a general relation valid for heterogeneous
equilibria.

%%%%%%%%%%%%%%%%%%%%%%%%%%%%%%%%%%%%%%%%%%%%%%%%%%%%%%%%%%%%%%%%%%%%%%%%%%%%%%%
\section{Soluble surfactant - special cases and extensions}
\mylab{sec:surfsol-special}
%%%%%%%%%%%%%%%%%%%%%%%%%%%%%%%%%%%%%%%%%%%%%%%%%%%%%%%%%%%%%%%%%%%%%%%%%%%%%%%
%
In this section we explore further the general gradient dynamics model
(\ref{eq:lubons})-(\ref{eq:psions}).
In particular,  we first show that well known hydrodynamic long-wave models are recovered as limiting cases. 
We then discuss extensions incorporating physical effects of interest that can be described within the present framework.
%
%%%%%%%%%%%%%%%%%%%%%%%%%%%%%%%%%%%%%%%%%%%%%%%%%%%%%%%%%%%%%%%%%%%%%%%%%%%%%%%
\subsection{Hydrodynamic formulation in dilute limit}
\mylab{sec:surfsol-class}
%%%%%%%%%%%%%%%%%%%%%%%%%%%%%%%%%%%%%%%%%%%%%%%%%%%%%%%%%%%%%%%%%%%%%%%%%%%%%%%
%
The standard hydrodynamic long-wave model employed for thin films with
a soluble surfactant that is dilute within the film and also has a low
coverage at the film surface \cite{OrDB1997rmp,CrMa2009rmp} is recovered from the general
gradient dynamics form (\ref{eq:lubons})-(\ref{eq:psions}) for zero
rigidity ($\kappa=0, \kappa_\mathrm{s}=0$), and with only the low-concentration
entropic (ideal-gas) terms in the energy -- i.e.\ neglecting the nonlinear
interaction terms in the
energies. Then, Eqs.~(\ref{eq:enfsurf0}) and ~(\ref{eq:enfbulk0}) become
\begin{equation}
 f_\mathrm{s} (\Gamma) = \gamma_0 + 
\frac{kT}{l_\mathrm{s}^2}\Gamma(\ln\Gamma -1), 
\qquad\mathrm{and}\qquad
f (\phi) = \frac{kT}{l^{3}}\phi(\ln\phi-1) 
\mylab{eq:enfsimp} \end{equation}
respectively, where 
$\gamma_0$ is a constant. The energy functional (\ref{eq:ff}) in the long-wave approximation is
\begin{equation}
\mathcal{F} \,=\,
 \int\left[  
h f(\phi) + f_\mathrm{s} (\Gamma) \xi + g(h)
\right] dx\,dy,
\mylab{eq:funclw}
\end{equation}
where $\xi=1+\frac{1}{2}(\nabla h)^2$.
Note that in (\ref{eq:funclw}) one has to write $\phi=\psi/h$ and
$\Gamma=\widetilde\Gamma/\xi$ to obtain the variations w.r.t.\ the
independent fields $h$, $\psi$ and $\widetilde\Gamma$, as discussed at the begin of section~\ref{sec:surfsol:energy}.
The variations are
\begin{eqnarray}
p = \frac{\delta \mathcal{F} }{\delta h} &=& -\partial_x(\gamma(\Gamma) \partial_x h) - \Pi(h) -\frac{kT}{l^{3}}\phi,\nonumber\\
\mu_\mathrm{s} = \frac{\delta \mathcal{F} }{\delta \widetilde\Gamma} &=& \frac{kT}{l^{2}_\mathrm{s}}\ln\Gamma,\nonumber\\
\mu = \frac{\delta \mathcal{F} }{\delta \psi} &=& \frac{kT}{l^{3}}\ln\phi,\mylab{eq:simpvari}
\end{eqnarray}
where $\gamma(\Gamma)=f_\mathrm{s}-\Gamma
f'_\mathrm{s}=\gamma_0-kT\Gamma/l^{2}_\mathrm{s} = \gamma_0-\gamma_\Gamma\Gamma$, i.e., purely
entropic low-concentration contributions to the free energy result in
a linear equation of state.  As a result, the evolution equations
(\ref{eq:lubons})-(\ref{eq:solsurf-mob-nc}) become
\begin{eqnarray}
\partial_t h &=&  \nabla \cdot
\left( \frac{h^3}{3 \eta}\nabla \left[-\nabla\cdot(\gamma \nabla h) - \Pi(h) \right] 
 + \frac{\gamma_\Gamma h^2}{2 \eta} \nabla\Gamma
 \right) \nonumber\\
&& - \beta_\mathrm{evap}\left( \hat\mu -\nabla\cdot(\gamma \nabla h) - \Pi(h) -\frac{kT}{l^{3}}\phi \right),
\mylab{eq:hsimp} \\
\partial_t \Gamma &=&  \nabla \cdot
\left( \frac{h^2\Gamma}{2 \eta}\nabla \left[-\nabla\cdot(\gamma \nabla h) - \Pi(h) \right]
  + \left(\frac{\gamma_\Gamma h\Gamma}{\eta} + D_\mathrm{s}\right) \nabla\Gamma
  \right) 
+ \frac{\beta}{l}\left(  \ln\phi - \ln\Gamma \right),
\mylab{eq:gamsimp}\\
\partial_t \psi &=&  \nabla  \cdot
\left( \frac{h^2\psi}{3 \eta}\nabla \left[-\nabla\cdot(\gamma \nabla h) - \Pi(h) \right]
  + \frac{\gamma_\Gamma h\psi}{2 \eta} \nabla\Gamma
+ D h \nabla \phi \right) 
- \beta\left(  \ln\phi - \ln\Gamma \right),
\mylab{eq:psisimp}
\end{eqnarray}
where we have assumed $l_\mathrm{s}=l$, 
used the mobility functions (\ref{eq:molmobs}) and
introduced $\beta=\beta_\mathrm{\psi\Gamma}kT/l^2$ and $\hat\mu=
-p_\mathrm{vap}$.  Note, that in the capillary terms
$\gamma=\gamma(\Gamma)$ is often replaced by $\gamma_0$ and that $\beta$ may still depend on
the concentrations. 

The model can be related to the standard hydrodynamic long-wave models
for films with soluble surfactants found in the literature.  In the
simple case without solvent evaporation ($\beta_\mathrm{evap} =0$ and
without wettability ($\Pi=0$), it corresponds to Eqs.~(117-119) of the
review \cite{CrMa2009rmp} if the expression $\ln\phi - \ln\Gamma $ in
our adsorption flux is replaced by the linearised $\phi-\Gamma$ as
already discussed in section~\ref{sec:surfsol-nonconsfluxadsorp}.
Eqs.~(21) of \cite{JeGr1993pfa} [also cf.~Eqs.~(4.29a-c) of the
review~\cite{OrDB1997rmp}] further neglect all Laplace pressure
contributions (equivalent to $\gamma\approx0$, but keeping Marangoni
flows) and adds permeability of the substrate for the surfactant.
In Ref.~\cite{WaCM2003jcis} the case of a volatile solvent is studied
for a surfactant-covered film on a heated substrate. Their
Eqs.~(50-52) add thermal Marangoni flows to our Eqs.~(\ref{eq:hsimp})
- (\ref{eq:psisimp}), have a linearised adsorption flux and an
evaporation flux $\sim 1/(h+K)$ that in our equation corresponds to
$\beta_\mathrm{evap}\sim 1/(h+K)$ and a $\hat\mu$ that is much
larger than the other evaporation terms. 
%
%%%%%%%%%%%%%%%%%%%%%%%%%%%%%%%%%%%%%%%%%%%%%%%%%%%%%%%%%%%%%%%%%%%%%%%%%%%%%%%
\subsection{Mixture of liquids without surfactant}
\mylab{sec:surfsol-marang}
%%%%%%%%%%%%%%%%%%%%%%%%%%%%%%%%%%%%%%%%%%%%%%%%%%%%%%%%%%%%%%%%%%%%%%%%%%%%%%%
%
Another important limit is the case of a liquid film of a binary mixture
that consists of components that change the surface tension without
forming a proper monolayer of surfactant molecules at the free
surface.
Refs.~\cite{Thie2011epjst,ThTL2013prl} presented a two-field gradient
dynamics model for the evolution of a film of a liquid binary mixture
on a solid substrate that allows for the description of coupled
dewetting and decomposition processes for arbitrary bulk (mixing)
energies including bulk rigidity terms, capillarity and wetting
energies that may depend on the film height and concentration. The two
fields are the film height $h$ and the effective solute layer height
$\psi$. The model recovers, for instance, the long-wave limit of
model-H (Navier-Stokes Cahn-Hilliard equations) as derived in
\cite{NaTh2010n}, but also goes far beyond as it allows for a number
of other systematic extensions \cite{Thie2011epjst,ThTL2013prl}.

However, this two-field model has an important shortcoming: in
Ref.~\cite{ThTL2013prl} it was noted that no obvious way exists to
incorporate a concentration-dependent surface tension into the model
without breaking the gradient dynamics structure. This implies that
introducing a Marangoni flow caused by the solutal Marangoni effect
into the hydrodynamic two-field thin-film model for a mixture could
break the thermodynamic consistency: If one incorporates a
concentration-dependent surface tension directly into the energy
functional [$\gamma(\phi)$ in Eq.~(1) of \cite{ThTL2013prl}] that only
depends on the height-averaged bulk concentration $\phi$ and film
height $h$, a Marangoni-like flux term is obtained, however, with the
wrong prefactor in the mobility function.  Therefore the use of the
model in Ref.~\cite{ThTL2013prl} is limited to cases where surface
activity can be neglected.

Here, in the context of the three-field model, this issue is resolved
in the following way. We show that one may take the full gradient
dynamics model for soluble surfactants introduced above in
Section~\ref{sec:surfsol:graddyn} and consider the limit of very fast
(instantaneous) adsorption/desorption. This limit corresponds to
$\beta_\mathrm{\psi\Gamma}\gg1$ in Eqs.~(\ref{eq:gamons})
and~(\ref{eq:psions}) implying that the non-conserved fluxes
equilibrate fast. As a result, on the slower time scale of the
conserved fluxes one has $J_\mathrm{ad}\approx0$
(cf.~Eq.~(\ref{eq:exflux1})-(\ref{eq:exflux3})) and the surfactant
concentration at the free surface is slaved to the one in the bulk
film. The dependence corresponds to the equilibrium relations
discussed in section~\ref{sec:surfsol-nonconsfluxadsorp}.

For example, in the case without rigidity one has
$f'(\phi)=f_\mathrm{s}'(\Gamma)/l_\mathrm{s}$ and in the limit of low
concentrations $\phi\ll1$ \textit{and} $\Gamma\ll1$ one obtains
$l_\mathrm{s}^3\ln\phi = l^3\ln\Gamma$ implying
$\Gamma=\phi^{(l_\mathrm{s}/l)^3}$. For
$l_\mathrm{s}=(1+\varepsilon) l$ and $\varepsilon\ll1$,
$\Gamma=\phi +3\varepsilon\phi\ln\phi
+O(\varepsilon^2)$. 
Assuming $\Gamma\approx\phi$ (i.e., $l_\mathrm{s}\approx l$, the
governing equations (\ref{eq:lubons})-(\ref{eq:solsurf-mob}) with the
mobility functions (\ref{eq:molmobs}), can be simplified by
multiplying Eq.~(\ref{eq:gamons}) by $l$ and adding it to
Eq.~(\ref{eq:psions}). As a result, an evolution equation for
$\widetilde\psi=\psi + l\Gamma=(h+l) \phi \approx h\phi=\psi$ is
obtained where we use $h\gg l$. Dropping the tilde and approximating
the mobilities according to $h\gg l$, the equation reads
\begin{equation}
\partial_t \psi \,=\,
\nabla\cdot\left[
\frac{h^2 \psi}{3 \eta}  \nabla\frac{\delta F}{\delta
  h} \,+\,
\left(\frac{\psi^2}{2 \eta} + \frac{D_\mathrm{s}l^3\phi}{kT}\right)
\nabla\frac{\delta F}{\delta \widetilde\Gamma}
\,+\,
\left(\frac{h\psi^2}{3 \eta} + \frac{D l^3\psi}{kT}\right)
\nabla\frac{\delta F}{\delta \psi}
\right].
\mylab{eq:solsurf-coup-grad-psi-maran}
\end{equation}
The film height equation (\ref{eq:lubons}) becomes
\begin{equation}
\partial_t h \,=\,
\nabla\cdot\left[\frac{h^3}{3\eta}\nabla\frac{\delta F}{\delta h}
\,+\,\frac{h^2\phi}{2\eta} \nabla\frac{\delta F}{\delta \widetilde\Gamma}
\,+\frac{h^2\psi}{3\eta}\nabla\frac{\delta F}{\delta \psi}
\right] -\,\beta_\mathrm{evap}
\left(\frac{\delta F}{\delta h} -p_\mathrm{vap}\right).
\mylab{eq:solsurf-coup-grad-h-maran}
\end{equation}

As we are in the dilute limit for $f_\mathrm{s}$, the second term in
the conserved part of (\ref{eq:solsurf-coup-grad-h-maran}) becomes
$\gamma_\Gamma h^2 \nabla \phi /2$ with $\gamma_\Gamma=kT/l^2$
corresponding to the standard form of the Marangoni flux.  The
hydrodynamic form of Eq.~(\ref{eq:solsurf-coup-grad-h-maran}) is then
\begin{equation}
\partial_t h \,=\,
\nabla\cdot\left[-\frac{h^3}{3\eta}\nabla\left(
\gamma_0\Delta h 
+ \Pi(h)  \right)
\,+\,\gamma_\Gamma\frac{h^2}{2\eta} \nabla \phi\right]
\mylab{eq:solsurf-coup-grad-h-maran2}
\end{equation}
while Eq.~(\ref{eq:solsurf-coup-grad-psi-maran}) becomes (again with
$h\gg l$ and approximating $\gamma$ by the reference value $\gamma_0$ in the capillary term)
\begin{equation}
\partial_t (\phi h) \,=\,
\nabla\cdot\left[-\frac{h^3\phi}{3\eta}\nabla\left(
\gamma_0\Delta h 
+ \Pi(h) \right)
\,+\,\left(\gamma_\Gamma\frac{h^2\phi}{2\eta} \,+\,D h \right)
 \nabla \phi \right]
\mylab{eq:solsurf-coup-grad-psi-maran2}
\end{equation}
with the bulk diffusion constant
$D$. Eqs.~(\ref{eq:solsurf-coup-grad-h-maran2}) and
(\ref{eq:solsurf-coup-grad-psi-maran2}) correspond exactly to the
hydrodynamic thin film equations employed, e.g., in the study of
coalescence and non-coalescence of sessile drops of mixtures in
Ref.~\cite{BMBK2011epje,KaRi2014jfm}. We emphasise that as shown
here they may be derived from the full three-field gradient dynamics model
in the dilute
limit. Remarkably, the resulting model can not be brought into the
form of a two-field gradient dynamics. This poses the intriguing
question whether there exist circumstances (consistent with the
employed approximations) where the broken gradient dynamics structure
can result in unphysical behaviour. This merits further consideration.
We finally remark that the proposed reduction from the three-field
gradient dynamics model to a two-field model also works for other
choices of the energies (also with rigidities) -- they only have to be
consistent between bulk and surface.

%%%%%%%%%%%%%%%%%%%%%%%%%%%%%%%%%%%%%%%%%%%%%%%%%%%%%%%%%%%%%%%%%%%%%%%%%%%%%%%
\subsection{Nonlinear equation of state}
\mylab{sec:surfsol-extensions-nonlin-eos}
%%%%%%%%%%%%%%%%%%%%%%%%%%%%%%%%%%%%%%%%%%%%%%%%%%%%%%%%%%%%%%%%%%%%%%%%%%%%%%%
%
In the literature, thin film dynamics is sometimes studied in the case
of soluble surfactants with equations similar to Eqs.~(\ref{eq:hsimp})
to (\ref{eq:psisimp}) but employing nonlinear equations of state
$\gamma(\Gamma)$ (e.g., Eqs.~(8)-(12) of
Ref.~\cite{WaCM2004pf}). Other examples of nonlinear equations of
state in thin film hydrodynamics are found in
Refs.~\cite{BoGr1988jfm,GaGr1990jfm,MaCr2001pf,HaSD2012sm}.  Often, the
nonlinearity is incorporated into the Marangoni term and the remaining
equation is left unchanged. This may lead to spurious results if the
underlying gradient dynamics structure is broken \cite{ThAP2016note2}. 
If instead, the free energy functional
is appropriately changed one finds that Marangoni flux, diffusion and
adsorption/desorption terms all change in a consistent manner.

In the case without rigidity 
($\kappa=\kappa_\mathrm{s} =0$) and without evaporation the resulting equations are
\begin{equation}
\partial_t h = -\nabla \cdot \left[\frac {h^3}{3 \eta}\nabla (\nabla \cdot \left(\gamma(\Gamma) \nabla h \right) 
+ \Pi(h) ) 
 + \frac{h^2}{2 \eta}\nabla \gamma(\Gamma) \right] 
\mylab{eq:lub3}
\end{equation}
\begin{eqnarray}
\partial_t \Gamma &=&  -\nabla \cdot
\left\{ \frac {h^2 \Gamma}{2\eta}\nabla (\nabla \cdot \left(\gamma(\Gamma) \nabla h \right) 
+\Pi(h) )
 +  \left[\frac {h \Gamma}{\eta} + \frac{D_\mathrm{s} l_\mathrm{s}^2}{kT}\right] \nabla \gamma(\Gamma) \right\}
 + \widetilde{J}_\mathrm{ad}(\Gamma,\phi),
\mylab{eq:gameq3}\\
\partial_t \psi &=&   -\nabla  \cdot
\left\{ \frac {h^2 \psi}{3\eta} \nabla (\nabla \cdot \left(\gamma(\Gamma) \nabla h \right) 
+ \Pi(h) ) 
 + \frac{h}{2\eta} \psi \nabla \gamma(\Gamma) + \frac{D l^3 h}{kT} \nabla p_\mathrm{osm}(\phi) \right\}
 - J_\mathrm{ad}(\Gamma,\phi),
\mylab{eq:psieq3}
\end{eqnarray}
where we used $\Gamma\nabla f'_s(\Gamma) = -\nabla \gamma(\Gamma)$ and
$\phi\nabla f'(\phi) = -\nabla p_\mathrm{osm}$
to express surface and bulk diffusion in terms of the surface tension
and osmotic pressure, respectively. For a discussion of the adsorption
fluxes see section~\ref{sec:surfsol-nonconsfluxadsorp}.

Nonlinear equations of state used in the literature are, for instance, the Scheludko equation of state \cite{BoGr1988jfm,GaGr1990jfm,WaCM2004pf}
\begin{equation}
\gamma(\Gamma) = \frac{\gamma_0}{[1+\theta\Gamma]^3};  
\mylab{eq:eos:scheludko}
\end{equation}
the exponential relation $\gamma(\Gamma) = \exp(-\alpha\Gamma)$
\cite{MaCr2001pf}; and the expression
$\gamma(\Gamma) = \gamma_0 -
RT\Gamma_\infty\ln(1-\Gamma/\Gamma_\infty)$
\cite{HaSD2012sm}.  If diffusion is expressed in the form of Fick's
law $j_\mathrm{diff}=\widetilde{D}(\Gamma)\nabla\Gamma$, the nonlinear
`diffusion constant' $\widetilde{D}(\Gamma) $ should then be
proportional to $d\gamma(\Gamma)/d\Gamma$ -- if a constant molecular
diffusivity $D_\mathrm{s}$ is assumed, cf.~Eq.~(\ref{eq:gameq3}). If
one does not assume
$\widetilde{D}(\Gamma)\sim d\gamma(\Gamma)/d\Gamma\sim -\Gamma
f''_s(\Gamma)$, as is the case in all the mentioned works, then it should
be realised
that implicitly a certain nonlinear dependence of the
molecular diffusivity on the concentration is being assumed,
that may often not be justified.
%
%%%%%%%%%%%%%%%%%%%%%%%%%%%%%%%%%%%%%%%%%%%%%%%%%%%%%%%%%%%%%%%%%%%%%%%%%%%%%%%
\subsection{Concentration-dependent wettability}
\mylab{sec:surfsol-extensions-wett}
%%%%%%%%%%%%%%%%%%%%%%%%%%%%%%%%%%%%%%%%%%%%%%%%%%%%%%%%%%%%%%%%%%%%%%%%%%%%%%%
%
The energy functional $\mathcal{F}$ described above in
section~\ref{sec:surfsol:energy} contains well separated bulk
contributions $\mathcal{L}_\mathrm{b}$ and surface contributions
$\mathcal{L}_\mathrm{s}$, namely Eqs.~(\ref{eq:solsurf}) and
(\ref{eq:insolsurf}), respectively. Energetic couplings (terms that
depend on more than one of the independent fields) exist due to the
surface metric and the introduction of the three independent fields
$h$, $\widetilde\Gamma$ and $\psi$. However, the bulk free energy
$f(\phi)$, surface free energy $f_\mathrm{s} (\Gamma)$ and wetting
energy $g(h)$ may also depend on the other fields. First, we discuss a
concentration-dependent wetting energy.

It has been discussed several times how to incorporate such a
dependency into the known hydrodynamic long-wave equations.
One approach is to make the interaction constants within the Derjaguin pressure
to depend on the surfactant concentration (case of insoluble surfactant)
\cite{WaCM2002pf,Hu2005pf,FiGo2007jcis,LiJY2013aps}. Another is to
make the (structural)
Derjaguin pressure to depend on the concentration of nanoparticles
to model layering effects \cite{Hu2012ams}.  Ref.~\cite{CrMa2007l}
includes a concentration-dependent disjoining pressure, and accounts
for surfactant layers at the free surface \textit{and} the solid substrate.
In the bulk film dissolved surfactant molecules as well
as micelles are considered. Similar extensions are made in Ref.~\cite{FiGo2007jcis}
for a two-layer system with surfactant.

We argue that incorporating such concentration-dependence
of wetting and dewetting phenomena has to start with an amended energy
functional. Then, a concentration-dependent Derjaguin pressure as
introduced in all the papers cited in the previous paragraph, is one
natural consequence \textit{but is not the only one}. We illustrate
this by replacing $g(h)$ in
Eq.~(\ref{eq:solsurf}) by the general expression
$g(h,\Gamma,\phi)$ for the case without rigidities
($\kappa=\kappa_s=0$) but keep $f=f(\phi)$ and
$f_\mathrm{s}=f_\mathrm{s}(\Gamma)$. Then the variations in long-wave
approximation are 
\begin{eqnarray}
p = \frac{\delta  F}{\delta h} \,&=&\, 
 f - \phi \partial_\phi f + \partial_h g - \frac{\phi}{h}\partial_\phi g
-\nabla\cdot\left(\tilde\omega\nabla h\right)
\mylab{eq:app-variwetta}\\
\mu_\mathrm{s} = \frac{\delta  F}{\delta \widetilde{\Gamma}}
\,&=&\,\partial_\Gamma g + \partial_\Gamma f_\mathrm{s}  
\mylab{eq:app-variwettb}\\
\mu = \frac{\delta  F}{\delta \psi}
\,&=&\,\frac{1}{h} \partial_\phi g + \partial_\phi f 
\mylab{eq:app-variwettc}
\end{eqnarray}
with the generalised surface tension
\begin{equation}
\tilde\omega=f_\mathrm{s}
-\Gamma\partial_\Gamma f_\mathrm{s} - \Gamma\partial_\Gamma g.
\mylab{eq:omegawett}
\end{equation}
Note the new contributions that depend on $\partial_\phi g$ or
$\partial_\Gamma g$ which appear in $p$, $\mu_\mathrm{s}$, $\mu$ and
$\tilde\omega$. They are often missing in the literature.
The full expressions for $\kappa\neq0$, $\kappa_s\neq0$ and general
$f$ and $f_\mathrm{s}$ are given in Appendix~\ref{app:variations}.

With Eqs.~(\ref{eq:app-variwetta}) to (\ref{eq:app-variwettc}) the general gradient dynamics form
(\ref{eq:lubons})-(\ref{eq:solsurf-mob}) of the evolution equations becomes
\begin{eqnarray}
\partial_t h &=&  \nabla\cdot  
\left( 
\frac{h^3}{3 \eta} 
\left[
\nabla \left( \partial_h g
- \nabla\cdot(\tilde\omega\nabla h)\right)
- \frac{\nabla \phi}{h}\partial_\phi g 
\right]\right.
\nonumber\\ 
&& \left.+ \frac{h^2\Gamma}{2 \eta}\nabla  
\left[
\partial_\Gamma g + \partial_\Gamma f_\mathrm{s}  
\right]\right) \,-\,  J_\mathrm{ev}(h,\Gamma,\phi),
\mylab{eq:weth} \\
\partial_t \Gamma &=&  \nabla\cdot
\left( \frac{h^2\Gamma}{2 \eta} 
\left[ \nabla \left( \partial_h g 
- \nabla\cdot(\tilde\omega\nabla h)\right)
- \frac{\nabla \phi}{h}\partial_\phi g
\right]
\right.
\nonumber\\ 
&& \left. 
  + \left(\frac{h\Gamma^2}{\eta} + \frac{D_\mathrm{s} l_\mathrm{s}^2 \Gamma}{kT}\right)\nabla  \left[
\partial_\Gamma g + \partial_\Gamma f_\mathrm{s}  
\right]\right) \,+\, \widetilde J_\mathrm{ad}(h,\Gamma,\phi),
\mylab{eq:wetgam}\\
\partial_t \psi &=&  \nabla\cdot  
\left( 
\frac{h^2\psi}{3 \eta}
\left[
\nabla\left( \partial_h g 
- \nabla\cdot(\tilde\omega\nabla h)\right)
- \frac{\nabla \phi}{h}\partial_\phi g
\right]
\right.
\nonumber\\ 
&& \left.  + \frac{h\psi\Gamma}{2 \eta}  \nabla  \left[
\partial_\Gamma g + \partial_\Gamma f_\mathrm{s}  
\right]
+ \frac{D l^3 \psi}{kT} \nabla 
\left[
\frac{1}{h}\partial_\phi g + \partial_\phi f 
\right]
\right) \,-\, l_\mathrm{s} \widetilde J_\mathrm{ad}(h,\Gamma,\phi).
\mylab{eq:wetpsi}
\end{eqnarray}
The non-conserved terms are only written in summary form, but can be
easily obtained with Eqs.~(\ref{eq:app-variwetta}) to (\ref{eq:app-variwettc})
from Eqs.~(\ref{eq:evapflux}) and (\ref{eq:exflux1}).

Inspecting Eqs.~(\ref{eq:weth})-(\ref{eq:wetpsi}), one notices that
the above mentioned cross-coupling terms depending on
$\partial_\phi g$ or $\partial_\Gamma g$ contribute to all conserved
and-non-conserved fluxes. These terms are important for very thin
films and in contact line regions where the free liquid-gas interface
approaches the solid-liquid interface. There they contribute to
diffusion, act as Marangoni-like driving terms of the
convective flux and influence adsorption and evaporation. For drops of
mixtures, a concentration-dependent wettability might, e.g., result in
a local phase decomposition in the contact line region or in a
single-component wetting layer (precursor films) as, e.g., observed in
experiments with polymer solutions \cite{FoBr1997el,FoBr1998m}. Note
that Derjaguin pressure isotherms for binary mixtures have already
been discussed in Ref.~\cite{DeCh1977jcis}.

It is our impression that the cross-coupling terms are often
missing in the literature. This is also important on general grounds since
without them the gradient dynamics structure of the dynamic equations
is broken. We believe that this is the reason why
Ref.~\cite{FiGo2007jcis} reports traveling and standing ``dewetting
waves'' that are clearly unphysical in a relaxational setting. It
seems also likely that the cusps in the dispersion curves obtained in
\cite{Hu2005pf} result from transitions between real and complex
eigenvalues. The latter could again result from a broken gradient
dynamics structure. However, the character of the eigenmodes is not
explicitly mentioned in Ref.~\cite{Hu2005pf}, here we only deduce this
possibility from the appearance of the dispersion curves.
%
%%%%%%%%%%%%%%%%%%%%%%%%%%%%%%%%%%%%%%%%%%%%%%%%%%%%%%%%%%%%%%%%%%%%%%%%%%%%%%%
\subsection{Surfactant phase transitions and mixture decomposition - bulk and surface rigidity}
\mylab{sec:surfsol-extensions-phase-trans}
%%%%%%%%%%%%%%%%%%%%%%%%%%%%%%%%%%%%%%%%%%%%%%%%%%%%%%%%%%%%%%%%%%%%%%%%%%%%%%%
%
In sections \ref{sec:surfsol-marang} and
\ref{sec:surfsol-extensions-nonlin-eos} we have discussed
concentration-dependent bulk energies $f(\phi)$ and surface energies
$f_\mathrm{s}(\phi)$. If these are nonlinear and exhibit negative second
derivatives, then the system is thermodynamically unstable over the
corresponding concentration range. In such a case a phase
decomposition in the bulk film \cite{GeKr2003pps} or a surfactant
phase transition \cite{RiSp1992tsf,LKGF2012s} may occur. Then a
theoretical description needs to include rigidity effects, i.e.,
$\kappa\neq0$ and/or $\kappa_s\neq0$ to assign an energetic cost to
strong concentration gradients. Long-wave models that include these
terms were already developed for non-surface active mixtures
\cite{NaTh2010n,ThTL2013prl} and non-soluble surfactants
\cite{KoGF2009el,KGFC2010l,ThAP2012pf}. In the case of constant
rigidities $\kappa_s$ and $\kappa$, a model for soluble surfactants
essentially combines the rigidity-related expressions developed in
\cite{ThTL2013prl} and \cite{ThAP2012pf}. Therefore, here we do not
explicitly write the bulky expressions. However, the variations of
the energy functional in the general case are given as
Eqs.~(\ref{eq:app-vari2a}) to (\ref{eq:app-vari2c}) in
Appendix~\ref{app:variations}, so the dynamic equations can be easily
obtained by introducing them into the general gradient dynamics form
(\ref{eq:lubons})-(\ref{eq:solsurf-mob}). The case of concentration
dependent rigidities may also be treated and these result in additional
contributions to the variations. Finally, note that the effect of
substrate-mediated condensation described in
\cite{RiSp1992tsf,LKGF2012s} naturally results in a free energy $f(\phi,h)$ that
depends on both $\phi$ and $h$, that is also covered in Appendix~\ref{app:variations}. 

This section ends the discussion of the special cases of the presented
general model. The following conclusion includes a discussion of
possible further extensions and open questions. Note, that there are
two appendices: Appendix~\ref{sec:lwvar} clarifies an issue in the
comparison of hydrodynamic long-wave approach and the present
variational approach and Appendix~\ref{app:variations} gives the
variations of the energy functional in the most general case covered
by the present work.
%
%%%%%%%%%%%%%%%%%%%%%%%%%%%%%%%%%%%%%%%%%%%%%%%%%%%%%%%%%%%%%%%%%%%%%%%%%%%%%%%
\section{Conclusions}
\mylab{sec:conc}
%%%%%%%%%%%%%%%%%%%%%%%%%%%%%%%%%%%%%%%%%%%%%%%%%%%%%%%%%%%%%%%%%%%%%%%%%%%%%%%
%
We have shown that a thin film (or long-wave) model for the dynamics
of liquid films on solid substrates with a free liquid-gas interface that
is covered by soluble surfactants can be brought into a gradient
dynamics form. Note that we always consider regimes where inertia does
not enter (small Reynold number).  The gradient dynamics form is fully
consistent with linear non-equilibrium thermodynamics including
Onsager's reciprocity relations \cite{Doi2011jpcm}. In the dilute
limit, the model reduces to the well-known hydrodynamic form that
includes Marangoni fluxes due to a linear equation of state relating
surface tension and surfactant concentration at the free surface
\cite{CrMa2009rmp}. In this case the free energy functional
incorporates wetting energy (resulting in a Derjaguin or disjoining
pressure), surface energy of the free interface (constant contribution
plus entropic term, resulting in capillarity - Laplace pressure - and
Marangoni flux) and bulk mixing free energy consisting solely of an
(ideal-gas) entropic term that results in a dependence of evaporation on osmotic
pressure but does not influence the convective flux. The entropic
contributions also determine surfactant diffusion within and on the
film and adsorption/desorption fluxes.

The advantage of the gradient dynamics form is that one may amend the
energy functional (incorporating non-entropic mixing and surface
energies, bulk and surface rigidities, concentration-dependent wetting
energies, etc.) and so one automatically obtains a thermodynamically
consistent set of updated expressions for the
Laplace and Derjaguin pressures, Marangoni,
Korteweg and diffusion fluxes, and evaporation as well as
adsorption/desorption terms. There are also new cross-coupling terms,
e.g., in the case of a concentration-dependent wettability. The
general model we have presented contains as limits the case of films of
non-surface active mixtures \cite{Thie2011epjst,ThTL2013prl} and
insoluble surfactants \cite{ThAP2012pf}. Such models with specific
energies are furthermore found in Refs.~\cite{NaTh2010n,STBT2015sm} and
\cite{KoGF2009el,KGFC2010l}, respectively. However, our work has also
shown that many models existing in the literature are incomplete because
they directly modify the hydrodynamic long-wave equations by
incorporating, e.g., concentration-dependent Derjaguin pressures or
nonlinear equations of state (for examples see
section~\ref{sec:surfsol-special}, but also the discussions in
\cite{Thie2011epjst,ThTL2013prl,ThAP2012pf}).  Such \textit{ad-hoc}
changes should be avoided as they alter only one `transport channel'
(e.g. Marangoni flux or pressure gradient) while the underlying change
of the energy functional affects all transport channels. So does, e.g.,
a change in the concentration-dependence of the surface free energy. This
not only changes the surface equation of state and the Marangoni flux,
but also affects surfactant diffusion and adsorption/desorption. A
concentration-dependent wettability results in a
concentration-dependent Derjaguin pressure and furthermore it gives a
new Marangoni-type flux, affects diffusion, evaporation, and
adsorption/desorption. We expect that our general model with
appropriately adapted energies can describe the film dynamics and incorporate
the effects of, e.g., the spreading of patches of high-concentration surfactants
on a liquid layer, that exhibit a local concentration maximum at the
advancing surfactant front \cite{FLFD2010njp,SHSG2014pf}, or the
adsorption/desorption dynamics of nanoparticles that act as surfactant
\cite{Bink2002cocis,GaCS2012l}.

Besides the amendments to the energy functional that we have discussed
at length, an important element of a thermodynamically-consistent
gradient dynamics structure are the mobilities that form a
positive-definite (positive entropy production) and symmetric
(Onsager's reciprocity relations) matrix. Whenever a similar model for
a relaxational situation is derived by making a long-wave approximation, a
transformation into the gradient dynamics form should result in such a
mobility matrix - thereby providing a valuable check that not all
models in the literature pass. Here, we have not changed the
convective mobilities, but allowed for general diffusive ones,
$M(\phi)$ and $M_\mathrm{s}(\Gamma)$. A further discussion of the
former [$M(\phi)$] is found in \cite{XuTQ2015jpcm}, where a
solvent-solute symmetric model is developed (without surface activity)
that is valid also for high solute concentrations. However, the
convective mobilities may also be amended: for instance, one can
incorporate slip at the substrate or solvent diffusion along the
substrate as discussed in Refs.~\cite{MWW2005jem} and \cite{HLHT2015l}
for films of simple liquids and layers of organic molecules,
respectively. Less is known about the mobility coefficients of the non-conserved
fluxes, so they are often approximated as a constant. A discussion of
different mobility functions in the evaporation term is found,
e.g., in \cite{Thie2014acis}, although there also a constant is
often used \cite{LyGP2002pre}. The influence of the mobilities should be
further studied -- in the present three-field case we expect a larger
influence than in the one-field case of a film of simple
liquid. There, the various convective mobilities mainly change the
relative timing of the different stages of the time evolution without
much change to the pathway itself \cite{HLHT2015l}.  Another important
factor that we have not discussed here, is the dependence of the liquid viscosity
on solute concentration. This is easy to incorporate, as long as the
liquid is Newtonian. A further future task is the incorporation of
surface viscosity \cite{SDDR2010el} that should results in changes to
the mobility matrix.

The gradient dynamics approach that we have presented may also be applied to
situations where more than the three fields considered here (effective
bulk solute height, projected surface concentration, film height)
matter. For example, systems with surfactant adsorption at the
solid substrate have relevance, e.g, for chemically-driven running
droplets \cite{SNKN2005pre,SuMY2006ptps} where the transfer of a
surfactant between different media and a solid substrate plays an
important role. To model such systems one needs to account for
adsorption at the substrate and diffusion of the adsorbate along the
substrate. This can be achieved through the incorporation of a fourth
field (adsorbate concentration) into the gradient dynamics structure
and an appropriate amendment of the energy functional. This leads to a
fourth evolution equation that couples through additional
adsorption/desorption fluxes with the dynamics of the other fields.
Such considerations are also important if one is seeking to model the
dependence of the fluid dynamics in the contact line region on the concentration,
including the concentration-dependence of all the involved interfacial
tensions and of the equilibrium contact angle. Such a model would allow one to
describe the dynamics of effects like, e.g., surfactant-induced
autophobing \cite{BDSE2016sm}.

Another important extension is the incorporation of micelle dynamics
\cite{CrMa2006pf,EdCM2006jfm}. This plays an important role, e.g., for
super-spreading, as does adsorption at the substrate
\cite{KaCM2011jfm,NiWa2011epjt,Mald2011jfm}. To do this, one must again
incorporate additional fields into the gradient dynamics approach. One
could employ the free energy approach of Ref.~\cite{HaDA2011jpcb} and
combine it with the present ideas to obtain coupled equations for the film
height, effective solute height, effective micellar height and surface
concentrations. This is straightforward if the
micelles are monodisperse in size. However, the number of equations will
proliferate if the number of molecules per micelle is considered in
detail.  In hydrodynamic long-wave models only one size is normally
considered \cite{EdCM2006jfm,BeMC2009l,CrMa2009rmp}.

Since the adsorption at the substrate may be physisorption or
chemisorption, the question arises whether, in general, chemical
reactions may be incorporated into a gradient
dynamics. Ref.~\cite{Miel2011n} provides such a formulation for
reaction-diffusion systems that may be coupled to the present
formulation of thin film hydrodynamics. Preliminary considerations
show that this is possible and results, e.g., in cross-couplings
between chemical reactions and wettability. However, as briefly
discussed in Section~\ref{sec:surfsol-nonconsfluxadsorp}, what the
correct way to construct the mobilities such that they agree with the
ones obtained via kinetic considerations is still an open question.

% Membrane adhesion at solid substrates is sometimes studied with thin
% film models including surfactant-like lipids and receptors that may
% bind to the solid substrate in a chemical reaction. A three variable
% model is presented in section~4.2 of the review
% \cite{GaCo1996hcr}. There the dynamics of film height, surfactant
% concentration and the concentration of molecules bound to the
% substrate are modeled through a combination of elements from thin-film
% theory and reaction-diffusion systems. Such systems can now be
% considered in a 'universal' gradient dynamics approach.
%

Throughout the present work we have nearly exclusively referred to
relaxational situations, i.e., experimental settings without any
imposed influxes or through-flows of energy or mass, where the initial
state relaxes towards a minimum of the underlying energy functional.
However, the resulting gradient dynamics formulation for the time evolution
can now be supplemented by well-defined (normally non-variational)
terms to describe systems that are permanently out of equilibrium.
Example of this are film flows and drop dynamics on inclined planes where a
gradient dynamics model is obtained by incorporating the potential
energy of the liquid into the energy functional \cite{EWGT2016arxiv}.

Other examples include models for dip-coating and
Langmuir-Blodgett transfer processes where a film of solution or
suspension is transfered from a bath onto a moving plate
\cite{WTGK2015mmnp}. Then the relaxational gradient dynamics is
supplemented by a dragging or comoving frame term that together with
lateral boundary conditions representing the bath and the deposited layer,
respectively, effectively transforms the model into a non-relaxational
out-of-equilibrium model that often shows multistability or
self-organised pattern formation
\cite{KGFC2010l,KGFT2012njp,KoTh2014n,WTGK2015mmnp}.  It is similar
for dragged films of simple liquids (aka the Landau-Levich problem)
\cite{SZAF2008prl,GTLT2014prl}, films and drops on/in rotating
cylinders \cite{Moff1977jdm,LRTT2016pf} and also for evaporative
dewetting of suspensions (in the comoving frame of a planar
evaporation front) \cite{frat2012sm,Thie2014acis}. Furthermore, one
may impose certain in- and/or out-fluxes of material that break the
gradient dynamics structure (e.g., caused by heating) \cite{BeMe2006prl}.

Finally, we point out that such an approach to interface-dominated
out-of-equilibrium processes may also be applied to the modelling of
(bio-)active soft matter. For instance, Ref.~\cite{TrJT2016ams}
presents a model for the osmotic spreading dynamics of bacterial
biofilms where a relaxational model for a mixture of aqueous solvent
and biomass is supplemented by growth terms that model the
proliferation of biomass. Another example considers a dilute carpet of
insoluble self-propelled micro-swimmers on a liquid film and describes
it using an extension of models developed for insoluble non-self-propelling
surfactant particles \cite{AlMi2009pre,PoTS2016epje}. To describe higher
concentrations of the micro-swimmers one could employ the present model
of soluble surfactants and add contributions resulting from the self-propulsion.

\acknowledgments

We acknowledge discussions with many colleagues about the concept of
gradient dynamics in the context of long-wave hydrodynamic models, for
instance, Richard Craster, Oliver Jensen, Michael Shearer and Tiezheng
Qian. We thank the Center of Nonlinear Science (CeNoS) and the
Internationalisation Funds of the Westf\"alische Wilhelms
Universit\"at M\"unster for their support of our collaborative
meetings and an extensive stay of LMP at M\"unster,
respectively. Further we would like to thank the Isaac Newton
Institute for Mathematical Sciences at the University of Cambridge for
the Research Program ``Mathematical Modelling and Analysis of Complex
Fluids and Active Media in Evolving Domains'' (2013) where where many
discussion with colleagues took place and the first part of this work
was perceived. We are thankful to Sarah Trinschek and Walter Tewes for
triple-checking part of our
calculations. % UT also acknowledges discussions with Ofer Manor in the
% final stage of the project and GIF grant no.\ I-1361-401.10/2016 for
% partial funding.

%%%%%%%%%%%%%%%%%%%%%%%%%%%%%%%%%%%%%%%%%%%%%%%%%%%%%%%%%%%%%%%%%%%%%%%%%%%%%%%
\newpage
\appendix
%%%%%%%%%%%%%%%%%%%%%%%%%%%%%%%%%%%%%%%%%%%%%%%%%%%%%%%%%%%%%%%%%%%%%%%%%%%%%%%
\section{Asymptotic long-wave expansion vs.\ variational approach}
\mylab{sec:lwvar}
%%%%%%%%%%%%%%%%%%%%%%%%%%%%%%%%%%%%%%%%%%%%%%%%%%%%%%%%%%%%%%%%%%%%%%%%%%%%%%%
%
There is an interesting issue in the variational form of the evolution
equations for an insoluble layer of surfactant on a liquid layer as
presented in Ref.~\cite{ThAP2012pf}. There, in Eq.~(15) the Laplace
pressure takes the form $-\partial_x (\gamma\partial_x h)$, where
$\gamma=\gamma(\Gamma)$ is the surfactant concentration-dependent
surface tension that emerges as the local grand potential
\cite{ThAP2016note3}.

Consider the curve representing the surface of a fluid in two
dimensions with surface tension $\gamma=\gamma(s)$ as a function of
arclength $s$.
On mechanical grounds one should expect that the force on a curve element to be the derivative
w.r.t.\ arclength of $\gamma(s)\vec{t}$, i.e.,
\begin{equation}
\frac{d}{ds}(\gamma(s)\vec{t}) = \frac{d\gamma(s)}{ds}\vec{t} +
\gamma(s)\frac{d\vec{t}}{ds} = \frac{d\gamma(s)}{ds}\vec{t} +
\gamma(s) K \vec{n}
\mylab{eq:curv}
\end{equation}
where
\[\vec{n}=\frac{1}{\xi}(-\partial_xh,1)^T,
\quad\vec{t}=\frac{1}{\xi}(1,\partial_xh)^T,
\quad K=\frac{\partial_{xx}h}{\xi^3}\]
are the normal vector, tangent vector and curvature of the surface, 
respectively, and $\xi=\bigl(1+(\partial_xh)^2\bigr)^{1/2}$.

This seems to indicate that the Laplace pressure term in a long-wave
model should be
$-\gamma\partial_{xx} h$ since Eq.~(\ref{eq:curv}) gives the r.h.s.\ of the
classical hydrodynamic force boundary condition (BC) at a free surface while the left hand
side is $(\vecg{\tau}_{in}-\vecg{\tau}_{out})\cdot\vec{n}$.

We show next that the form  $-\partial_x (\gamma\partial_x h)$ in Ref.~\cite{ThAP2012pf} that also
appears in all the models presented here naturally arises
when projecting the force BC not onto $\vec{n}$ and $\vec{t}$ (as
done for general interfaces), but onto the cartesian unit vectors $\vec{e}_x=(1,0)^T$
and $\vec{e}_z=(0,1)^T$, as appropriate when performing a long-wave approximation.

The stress tensor is
\begin{equation}
\vecg{\tau}\,=\,-p \vec{I} \,+\, \eta(\nabla\vec{v}+(\nabla\vec{v})^T).
\mylab{eq:stresstensor}
\end{equation}
where $p(x,z)$ stands for the pressure field and $\vec{I}$ is the identity tensor.
The force equilibrium is
\begin{equation} 
(\vecg{\tau}-\vecg{\tau}_\mathrm{air})\cdot\vec{n}\,=\,\gamma K\,\vec{n}\,+\,(\partial_\mathrm{s}\gamma)\,\vec{t}
\mylab{eq:bcforce}
\end{equation}
where the surface derivative is defined by $\partial_\mathrm{s}=\vec{t}\cdot\nabla$ and we
assume that the ambient air does not transmit any shear stress
($\vecg{\tau}_\mathrm{air}=p_\mathrm{gas}\vec{I}$) and introduce $p=p_\mathrm{liq}-p_\mathrm{gas}$.

The boundary condition (\ref{eq:bcforce}) is of vectorial character,
i.e.\ one can derive two scalar conditions by projecting it onto two
different directions.  In
Refs.~\cite{OrDB1997rmp,Thie2007,CrMa2009rmp} projections onto $\vec{n}$ and $\vec{t}$
are used, resulting in
\begin{align}
\vec{t}\;&:\quad
\eta\,[(u_z+w_x)(1-h_x^2)+2(w_z-u_x)h_x]=\partial_\mathrm{s}\gamma(1+h_x^2)
\mylab{eq-bcforce-proj1a}\\
\vec{n}\;&:\quad
           p+\frac{2\eta}{1+h_x^2}\bigl[-u_xh_x^2-w_z+h_x(u_z+w_x)\bigr]=-\gamma K
\mylab{eq-bcforce-proj1b} 
\end{align}
Note that to highest order in long-wave scaling (see below) this
results in BC (when keeping all the surface tension terms)
$p=-\varepsilon^2\gamma h_{xx}$ and $\eta
u_z=\varepsilon\partial_x\gamma$.

Here, instead, we project onto $\vec{e}_x$
and $\vec{e}_z$ obtaining
\begin{align}
\vec{e}_x\;&:\quad -h_x (2\eta u_x - p) + \eta\,(u_z + w_x) = - h_x \gamma
K +\partial_\mathrm{s}\gamma
\mylab{eq-bcforce-proj2a}\\
\vec{e}_z\;&:\quad  -\eta h_x (w_x+u_z) + 2\eta w_z - p = \gamma 
K + h_x \partial_\mathrm{s}\gamma
\mylab{eq-bcforce-proj2b} 
\end{align}
Next we introduce the long-wave scaling with length scale
ratio $\varepsilon =H/L$. Note, that we do not non-dimensionalize.
We also replace
$K\approx h_{xx}$ and $\partial_\mathrm{s}\gamma
\approx \partial_x\gamma$ - formally introducing
scaled (long-wave) variables $x'=\varepsilon x$ and $w'= w/\varepsilon$. After dropping the dashes
we have
\begin{align}
\vec{e}_x\;&:\quad -\varepsilon h_x (2\eta \varepsilon
u_x - p) + \eta\,(u_z + \varepsilon^2 w_x) = - \varepsilon^3 \gamma h_x 
h_{xx} +\varepsilon\partial_x\gamma
\mylab{eq-bcforce-proj3a}\\
\vec{e}_z\;&:\quad  -\varepsilon\eta h_x (\varepsilon^2 w_x+u_z) +
2\varepsilon \eta w_z - p = \varepsilon^2 \gamma 
h_{xx} + \varepsilon^2 h_x \partial_x\gamma
\mylab{eq-bcforce-proj3b} 
\end{align}
In the usual way \cite{Thie2007} one takes into account that all
velocities are small, introducing $u'=u/\varepsilon$,
$w'=w/\varepsilon$; 
dropping small terms with the exception of surface tension related
terms. After dropping the dashes one has
\begin{align}
\vec{e}_x\;&:\quad \varepsilon h_x p + \varepsilon \eta u_z = - \varepsilon^3 \gamma h_x 
h_{xx} +\varepsilon \partial_x\gamma
\mylab{eq-bcforce-proj4a}\\
\vec{e}_z\;&:\quad  - p = \varepsilon^2 \gamma 
h_{xx} + \varepsilon^2 h_x \partial_x\gamma
\mylab{eq-bcforce-proj4b} 
\end{align}
Introducing Eq.~(\ref{eq-bcforce-proj4b}) into
Eq.~(\ref{eq-bcforce-proj4a}) one has
\begin{equation} 
\varepsilon h_x (-\varepsilon^2 \gamma 
h_{xx} - \varepsilon^2 h_x \partial_x\gamma) + \varepsilon \eta u_z = - \varepsilon^3 \gamma h_x 
h_{xx} +\varepsilon \partial_x\gamma
\mylab{eq-bcforce-proj5a}
\end{equation}
i.e.\
\begin{equation} 
\eta u_z =  (1 + \varepsilon^2 h_x^2) \partial_x\gamma \approx \partial_x\gamma.
\mylab{eq-bcforce-proj5a}
\end{equation}
The second condition (\ref{eq-bcforce-proj4b}) is identical to 
\begin{equation} 
p = - \varepsilon^2 \partial_x (\gamma \partial_x h).
\mylab{eq-bcforce-proj5b}
\end{equation}
As the previous two equations give the BC for the bulk equations
$u_{zz}=p_x$ and $p_z=0$, the involved quantities have to scale as
$O(\varepsilon^2 \gamma)=O(\partial_x\gamma)=O(p)=O(u)=O(1)$, i.e.,
in other words
$\partial_x (\gamma \partial_x h)\approx \gamma \partial_{xx} h$.  The
difference is of higher order in $\varepsilon$. Our consideration
poses the interesting question whether an asymptotic expansion should
in general be done in such a way that it does not break deeper
principles.  Here the deeper principle is the thermodynamically
consistent gradient dynamics formulation required for the description
of a relaxational process. Therefore
$\partial_x (\gamma \partial_x h)$ should be preferred over
$\gamma \partial_{xx} h$.
%
%%%%%%%%%%%%%%%%%%%%%%%%%%%%%%%%%%%%%%%%%%%%%%%%%%%%%%%%%%%%%%%%%%%%%%%%%%%%%%%
\section{Variations in the general case}
\mylab{app:variations}
The free energy $F\left[h,\Gamma,\phi\right]$ for the thin liquid film covered with soluble surfactant (aka film of a mixture with surface active components) is 
\begin{equation}
F\left[h,\frac{\widetilde\Gamma}{\xi},\frac{\psi}{h}\right]\,=\,\int\left\{
h f\left(h,\frac{\psi}{h}\right)
+g\left(h,\frac{\widetilde\Gamma}{\xi},\frac{\psi}{h}\right) 
+\xi f_\mathrm{s}\left(h,\frac{\widetilde\Gamma}{\xi}\right)
+h \frac{\kappa}{2} \left(\nabla\frac{\psi}{h}\right)^2
+ \frac{\kappa_\mathrm{s}}{2} \frac{1}{\xi} \left(\nabla\frac{\widetilde\Gamma}{\xi}\right)^2
\right\}\,dA.
\mylab{eq:en:full}
\end{equation}
We define
\begin{equation}
F\left[h,\frac{\widetilde\Gamma}{\xi},\frac{\psi}{h}\right] = F_\mathrm{bulk} + F_\mathrm{wet} + F_\mathrm{surf} + F_\mathrm{gradbulk} + F_\mathrm{gradsurf}
\end{equation}
and separately calculate the variations of the five terms in the free energy.
For simplicity, we only consider the one-dimensional case. An extension to the general two-dimensional case is straightforward. 
Initially, we keep the full expression $\xi=\sqrt{1+(\partial_x h)^2}$
and introduce  the long-wave approximation for $\xi$ later on.
This implies
\begin{equation}
\frac{\partial}{\partial h} \xi =0,\quad
\frac{\partial \xi}{\partial(\partial_x h)}= \frac{1}{\xi}\partial_x h, \quad 
\partial_x \xi = \frac{1}{\xi}(\partial_x h)(\partial_{xx} h)
\quad\mbox{and}\quad
\frac{\partial}{\partial (\partial_x h)}  \frac{1}{\xi}  = -\frac{1}{\xi^3}\partial_x h.
\end{equation}
\subsection{Variations with respect to $h$}
\begin{equation}
\frac{\delta F_\mathrm{bulk}}{\delta h} = f + h \partial_h f - \phi \partial_\phi f
\end{equation}
\begin{equation}
\frac{\delta F_\mathrm{wet}}{\delta h} = \partial_h g - \frac{\phi}{h}\partial_\phi g +
\frac{d}{dx}\left[\frac{\Gamma}{\xi^2}(\partial_\Gamma g)\partial_x h\right]
\end{equation}
Note that the final term was missed in Eq.~(A4) of
  Ref.~\cite{ThAP2012pf}. This then also results in amendments in their
Eq.~(23), namely there is an additional $-\Gamma\partial_\Gamma g$ in
the surface tension $\gamma$ in their Eq.~(23) and the Marangoni force is
$\nabla\gamma - (\partial_\Gamma g)\nabla\Gamma$ (Note that our $g$ is
their $f$).

Next, we have
\begin{eqnarray}
\frac{\delta F_\mathrm{surf}}{\delta h} &=& \xi \partial_h f_\mathrm{s} - \frac{d}{dx}\left[
\frac{1}{\xi} f_\mathrm{s} \partial_x h -\frac{1}{\xi^2} (\partial_\Gamma f_\mathrm{s})\widetilde\Gamma\partial_x h
\right]\\
&=& \xi\partial_h f_\mathrm{s} - \frac{d}{dx}\left[\frac{1}{\xi}(f_\mathrm{s}
-\Gamma\partial_\Gamma f_\mathrm{s})\partial_x h\right].
\end{eqnarray}
For the next variation we need to use
\begin{equation}
\frac{\delta (\int\star dx)}{\delta h}= \frac{\partial \star}{\partial
  h}-\frac{d}{dx}\frac{\partial \star}{\partial (\partial_x h)}
+\frac{d^2}{dx^2}\frac{\partial \star}{\partial (\partial_{xx} h)}.
\end{equation}
We also need
\begin{eqnarray}
\partial_x
\frac{\widetilde\Gamma}{\xi}&=&\frac{\partial_x\widetilde\Gamma}{\xi}-\frac{\widetilde\Gamma}{\xi^2}\partial_x\xi\\
&=&\frac{\partial_x\widetilde\Gamma}{\xi}-\frac{\widetilde\Gamma}{\xi^3}(\partial_x h)(\partial_{xx} h)
\end{eqnarray}
The variations of the gradient terms are then
\begin{eqnarray}
  \frac{\delta F_\mathrm{gradbulk}}{\delta h} &=& 
\frac{\kappa}{2} \left(\partial_x\frac{\psi}{h}\right)^2
+\kappa \left(\partial_x\frac{\psi}{h}\right)  \left[ -\frac{\partial_x \psi}{h}+\frac{2\psi}{h^2}\partial_xh\right]
+\frac{d}{dx}\left[\kappa \frac{\psi }{h}  \left(\partial_x\frac{\psi}{h}\right)\right]
\nonumber\\
&=& 
\frac{\kappa}{2} \left(\partial_x\phi\right)^2
+\kappa \frac{\phi}{h}(\partial_x h)\left(\partial_x\phi\right)
+\kappa \phi  \partial_{xx}\phi
\end{eqnarray}
and 
\begin{eqnarray}
  \frac{\delta F_\mathrm{gradsurf}}{\delta h} &=& -\frac{d}{dx}\left[ 
-\frac{\kappa_\mathrm{s}}{2}\left(\partial_x\frac{\widetilde\Gamma}{\xi}\right)^2\frac{\partial_x h}{\xi^3}
-\frac{\kappa_\mathrm{s}}{\xi^4}\left(\partial_x\widetilde\Gamma\partial_x h
+\widetilde\Gamma\partial_{xx} h
-3\frac{\widetilde\Gamma}{\xi^2}(\partial_x h)^2\partial_{xx} h\right)\partial_x\frac{\widetilde\Gamma}{\xi}
\right]\nonumber\\
&&-\frac{d^2}{dx^2}\left[\frac{\kappa_\mathrm{s}}{\xi^4} \left(\partial_x\frac{\widetilde\Gamma}{\xi}\right)\widetilde\Gamma\partial_x h
\right]\nonumber\\
 &=& -\frac{d}{dx}\left\{ \frac{\kappa_\mathrm{s}}{\xi^3}\left[
-\frac{1}{2}\left(\partial_x \Gamma\right)^2\partial_x h
-\left(\partial_x\Gamma\partial_x h +\Gamma\partial_{xx} h
-2\frac{\Gamma}{\xi^2}(\partial_x h)^2\partial_{xx} h\right)\partial_x\Gamma
\right.\right.\nonumber\\
&&\left.\left.-\left(3\frac{\Gamma}{\xi^2}  (\partial_x  h)^2\partial_{xx} h
-  \partial_x\Gamma\partial_x h
-  \Gamma\partial_{xx} h\right)\partial_x\Gamma
+ \Gamma\partial_x h\partial_{xx}\Gamma
\right]\right\}\nonumber\\
 &=& \frac{d}{dx}\left\{ \frac{\kappa_\mathrm{s}}{\xi^3}\left[
\frac{1}{2}\left(\partial_x \Gamma\right)^2\partial_x h
+\frac{\Gamma}{\xi^2}(\partial_x h)^2(\partial_{xx} h)\partial_x\Gamma
- \Gamma\partial_x h\partial_{xx}\Gamma
\right]\right\}
\end{eqnarray}
\subsection{Variations with respect to $\widetilde\Gamma$}
\begin{equation}
\frac{\delta F_\mathrm{bulk}}{\delta\widetilde\Gamma} = 0 \quad\mbox{and}\quad \frac{\delta F_\mathrm{gradbulk}}{\delta\widetilde\Gamma} =0
\end{equation}
\begin{equation}
\frac{\delta F_\mathrm{wet}}{\delta\widetilde\Gamma} = \frac{1}{\xi}\partial_\Gamma g
\end{equation}
\begin{equation}
\frac{\delta F_\mathrm{surf}}{\delta\widetilde\Gamma} = \partial_\Gamma f_\mathrm{s}
\end{equation}
\begin{eqnarray}
  \frac{\delta F_\mathrm{gradsurf}}{\delta\widetilde\Gamma} &=&
-\kappa_\mathrm{s} \frac{1}{\xi^4} (\partial_x\Gamma) (\partial_x h)  (\partial_{xx} h) - \kappa_\mathrm{s}\frac{d}{dx}\left[
\frac{1}{\xi^2}\partial_{x}\Gamma
\right]\nonumber\\
&=&
\kappa_\mathrm{s} \frac{1}{\xi^4} (\partial_x\Gamma) (\partial_x h)  (\partial_{xx} h) - \kappa_\mathrm{s}
\frac{1}{\xi^2}\partial_{xx}\Gamma
\end{eqnarray}
\subsection{Variations with respect to $\psi$}
\begin{equation}
\frac{\delta F_\mathrm{surf}}{\delta\psi} = 0 \quad\mbox{and}\quad \frac{\delta F_\mathrm{gradsurf}}{\delta\psi} =0
\end{equation}
\begin{equation}
  \frac{\delta F_\mathrm{wet}}{\delta\psi} = \frac{1}{h} \partial_\phi g
\end{equation}
\begin{equation}
  \frac{\delta F_\mathrm{bulk}}{\delta\psi} = \partial_\phi f
\end{equation}
\begin{equation}
  \frac{\delta F_\mathrm{gradbulk}}{\delta\psi} =
-\kappa \frac{1}{h} (\partial_x\phi) (\partial_x h) 
- \kappa\partial_{xx}\phi
\end{equation}
\subsection{Collecting the terms}
The resulting expressions for the variations are
\begin{eqnarray}
p = \frac{\delta  F}{\delta h} \,&=&\, 
 f + h \partial_h f - \phi \partial_\phi f + \partial_h g - \frac{\phi}{h}\partial_\phi g + \xi\partial_h f_\mathrm{s}
\nonumber\\
&+&\frac{\kappa}{2} \left(\partial_x\phi\right)^2
+\kappa \frac{\phi}{h}(\partial_x h)\left(\partial_x\phi\right)
+\kappa \phi  \partial_{xx}\phi
\mylab{eq:app-vari1a}
\\
&-& \partial_{x}\left[\frac{1}{\xi}\left(f_\mathrm{s}
-\Gamma\partial_\Gamma f_\mathrm{s} - \frac{\Gamma}{\xi}\partial_\Gamma g
-\frac{\kappa_\mathrm{s}}{2\xi^2}\left(\partial_x \Gamma\right)^2
+ \frac{\kappa_\mathrm{s}}{\xi}\Gamma\partial_{x}\left(  \frac{1}{\xi}\partial_{x}  \Gamma\right)
\right)\partial_x h
\right]
\nonumber\\
\mu_\mathrm{s} = \frac{\delta  F}{\delta \widetilde{\Gamma}}
\,&=&\,\frac{1}{\xi}\partial_\Gamma g + \partial_\Gamma f_\mathrm{s}  
- \frac{\kappa_\mathrm{s}}{\xi}\partial_{x} \left(  \frac{1}{\xi}\partial_{x}\Gamma\right)
\mylab{eq:app-vari1b}
\\
\mu = \frac{\delta  F}{\delta \psi}
\,&=&\,\frac{1}{h} \partial_\phi g + \partial_\phi f 
- \frac{\kappa}{h} \partial_x(h \partial_x \phi) 
\mylab{eq:app-vari1c}
\end{eqnarray}
This seems the appropriate stage in the derivation to apply the long-wave
approximation, i.e., to use $(\partial_x
h)^2\sim\varepsilon^2\ll1$. Therefore $\xi\approx 1+O(\varepsilon^2)$
and one obtains to highest order
\begin{eqnarray}
p = \frac{\delta  F}{\delta h} \,&=&\, 
 f + h \partial_h f - \phi \partial_\phi f + \partial_h g - \frac{\phi}{h}\partial_\phi g + \partial_h f_\mathrm{s}
\nonumber\\
&+&\frac{\kappa}{2} \left(\partial_x\phi\right)^2
+\kappa \frac{\phi}{h}(\partial_x h)\left(\partial_x\phi\right)
+\kappa \phi  \partial_{xx}\phi
\mylab{eq:app-vari2a}\\
&-& \partial_x\left[\tilde\omega\partial_x h
\right]
\nonumber\\
\mu_\mathrm{s} = \frac{\delta  F}{\delta \widetilde{\Gamma}} 
\,&=&\,\partial_\Gamma (f_\mathrm{s}  + g)
- \kappa_\mathrm{s}\partial_{xx}\Gamma
\mylab{eq:app-vari2b}\\
\mu = \frac{\delta  F}{\delta \psi} 
\,&=&\, \partial_\phi f  + \frac{1}{h} \partial_\phi g 
- \frac{\kappa}{h} \partial_x(h \partial_x \phi) 
\mylab{eq:app-vari2c}
\end{eqnarray}
where we have introduced
\begin{equation}
\tilde\gamma =\tilde\omega=f_\mathrm{s}
-\Gamma\partial_\Gamma f_\mathrm{s} - \Gamma\partial_\Gamma g
-\frac{\kappa_\mathrm{s}}{2}\left(\partial_x \Gamma\right)^2
+ \kappa_\mathrm{s}\Gamma\partial_{xx}\Gamma
\mylab{eq:omegtilde}
\end{equation}
corresponding to the surface grand potential density for the nonlocal
case. Note that $\nabla \tilde\gamma=-\Gamma\nabla \mu_s - \partial_\Gamma\nabla\Gamma$.
The free energy in the general case (\ref{eq:en:full}) may be simplified by assuming that cross-couplings between composition
and film height are all contained in $g\left(h,\Gamma,\phi\right)$ and do not appear in the bulk and surface energy. The latter are then
$f\left(\phi\right)$ and $f_\mathrm{s}\left(\Gamma\right)$, respectively.
In consequence, $\partial_h f=0$ and $\partial_h f_\mathrm{s}=0$
Eqs.~(\ref{eq:app-vari2a})-(\ref{eq:app-vari2c}) simplify
accordingly.
The general expressions for the variations, i.e.,
Eqs.~(\ref{eq:app-vari2a}) to (\ref{eq:app-vari2c}) are then
introduced into the general gradient dynamics form
(\ref{eq:lubons})-(\ref{eq:solsurf-mob}). With specific simplifying
assumptions for the individual terms of the energy functional, one
obtains several models in the literature and all the models introduced above as
special cases.

%\bibliographystyle{unsrturl}
%\bibliography{ThAP2015}

\begin{thebibliography}{100}
\bibitem{ThAP2016note1}
Note, that formally since $\psi$ is an effective layer thickness, $\mu$ is also
  a pressure -- a partial pressure of the solute.
\bibitem{ThAP2016note2}
Normally, the equations can still be brought into the gradient dynamics form by
  suitably modifying the dependence of the molecular diffusivity on
  concentration.
\end{thebibliography}

\end{document}